\documentclass[12pt]{article}
\usepackage{amsmath}
\usepackage{amssymb}
\def\al{\alpha}

\def\ga{\gamma} 
\def\Ga{\Gamma}
\def\ep{\epsilon}

  \def\calL{{\cal L}}
  \def\calO{{\cal O}}

\def\del        {  \partial  }
\def\p          {\partial}
\def\Tr         { {\rm Tr} }
\def\ie         {  {\it i.e.}      }

\def\comma          {\, ,}
\def\period         {\, .}
\def\lsim    {\lower .65ex \hbox{\ $\stackrel{<}{\sim}$\ } }
\def\gsim    {\lower .65ex \hbox{\ $\stackrel{>}{\sim}$\ } }
\def\com#1#2   { \left[#1, #2\right]} 
\def\acom#1#2  {\left\{ #1,#2\right\}}
\def\bra#1     {\langle #1 |}
\def\ket#1     {| #1 \rangle}
 
\def\vecii#1#2      {  \left(\begin{array}{c}#1\\#2\end{array}\right)  }
\def\veciii#1#2#3   {  \left(\begin{array}{c}#1\\#2\\#3\end{array}
                     \right)  }
\def\veciv#1#2#3#4  {  \left(\begin{array}{c}#1\\#2\\#3\\#4
                                 \end{array}\right)  }
\def\vecfv#1#2#3#4#5 {  \left(\begin{array}{c}#1\\#2\\#3\\#4\\#5
                                 \end{array}\right)  }

\def\matrixii#1#2#3#4            {  \left(\begin{array}{cc}#1&#2\\#3&#4
                                       \end{array}\right) }
\def\matrixiii#1#2#3#4#5#6#7#8#9 {  \left(\begin{array}{ccc}#1&#2&#3\\
                                     #4&#5&#6\\#7&#8&#9\end{array}
                               \right)  }
\def\mativ#1#2#3#4               {  \left(\begin{array}{cccc}
                                       #1\\#2\\#3\\#4\end{array}\right) }
\def\matv#1#2#3#4#5              {  \left(\begin{array}{ccccc}
                                     #1\\#2\\#3\\#4\\#5\end{array}
                              \right)  }
\def\eqabegin         {  \begin{eqnarray}  }
\def\eqaend           {  \end{eqnarray}  }
\def\nn               {  \nonumber  }
\def\bracetwo#1#2     {  \left\{ \begin{array}{l} #1 \\ #2 \end{array}
                         \right.  }
\def\bracetwocases#1#2#3#4  {   \left\{ \begin{array}{ll} #1 &
                                 \qquad #2 \\
                                 #3 & \qquad #4 \end{array} \right.  }
\def\bracebegin#1     {  \left\{ \begin{array}{#1}   }
\def\braceend         {  \end{array}\right.   }
\def\parn              {  \par\noindent }

\def\parmedskip        {  \par\medskip  }
\def\parsmallskip      {  \par\smallskip  }
\def\parbigskipn        {  \par\bigskip\noindent  }
\def\parmedskipn        {  \par\medskip\noindent  }

\def\parag#1           {\paragraph{#1} \mbox{ }\parmedskip\noindent}
\def\msection#1      {  \begin{center} \section{#1} \end{center}   }
\def\nsection#1      {  \let\boldface\bf \def\bf{} \section{#1}
                           \let\bf\boldface   }
\def\mnsection#1     {  \begin{center} \nsection{#1} \end{center}  }
\def\capsection#1    {  \let\boldface\bf \def\bf{\sc} \section{#1}
                           \let\bf\boldface   }
\def\mcapsection#1   {  \begin{center} \capsection{#1} \end{center} }

\newcommand{\nullify}[1]{}
\def\papertitlepage{\baselineskip 3.5ex \thispagestyle{empty}}
\def\Title#1{\baselineskip 1cm \vspace{1.5cm}\begin{center}
 {\Large\bf #1} \end{center} 
\vspace{0.5cm}}
\def\Authors#1{\begin{center} {\it #1} \end{center}}
\def\Abstract{\vspace{1.0cm}\begin{center} {\large\bf Abstract} 
           \end{center} \par\bigskip}
\def\Komabanumber#1#2#3{\hfill \begin{minipage}{4.2cm} UT-Komaba #1
              \parn #2 
              \parn #3 \end{minipage}}

\renewenvironment{thebibliography}{\pagebreak[3]\par\vspace{0.6em}
\begin{flushleft}{\large \bf References}\end{flushleft}
\vspace{-1.0em}

\begin{enumerate}\if@twocolumn\baselineskip=0.6em\itemsep -0.2em
\else\itemsep -0.2em\fi\labelsep 0.1em}{\end{enumerate}}
\def\dirac#1{{\ooalign{\hfil/\hfil\crcr$#1$}}}

\def\shead#1{\parmedskipn {\large\bf #1}}
\def\and{&}
\def\brkeq{\nonumber \\[.8ex] &}
\def\landfz{\\[.3ex] }

\renewenvironment{thebibliography}{\pagebreak[3]\par\vspace{0.6em}
\begin{flushleft}{\large \bf References}\end{flushleft}
\vspace{-1.0em}

\begin{enumerate}\if@twocolumn\baselineskip=0.6em\itemsep -0.2em
\else\itemsep -0.2em\fi\labelsep 0.1em}{\end{enumerate} }
\numberwithin{equation}{section}
\numberwithin{figure}{section}
\numberwithin{table}{section}
\begin{document}
\papertitlepage
\vspace*{0cm}
\Komabanumber{01-02}{hep-th/0106218}{June, 2001}
\Title{Fully Off-shell Effective Action and its Supersymmetry
 in Matrix Theory II} 
\vspace{1cm}
\Authors{{\sc Y.~Kazama\footnote[2]{kazama@hep3.c.u-tokyo.ac.jp} 
 and T.~Muramatsu
\footnote[3]{tetsu@hep1.c.u-tokyo.ac.jp}
\\ }
\vskip 3ex
 Institute of Physics, University of Tokyo, \\
 Komaba, Meguro-ku, Tokyo 153-8902 Japan \\
  }
\baselineskip .7cm
\Abstract In a previous work, we computed the fully off-shell effective
action $\Gamma$ and the corresponding quantum-corrected supersymmetry
(SUSY) transformation operator $\delta_\epsilon$ for the so-called
source-probe configuration in Matrix theory at one loop at order 4 in
the derivative expansion, and showed that they satisfy the SUSY Ward
identity $\delta_\epsilon \Gamma=0$. In this article, starting from the
most general form of $\Gamma$, we demonstrate that, conversely, given
such $\delta_\epsilon$ the SUSY Ward identity determines $\Gamma$
uniquely to the order specified above.  Our demonstration does not
require the explicit knowledge of the quantum-corrected supersymmetry
transformation and hence strongly suggests that the uniqueness property
would persist to all orders in perturbation theory.
\newpage
\baselineskip 3.5ex
\section{Introduction}
This is a continuation of the previous investigations
 \cite{Kaz-Mura1,Kaz-Mura2} 
on the question of the power of supersymmetry (SUSY)
 in Matrix theory for M-theory \cite{bfss,susskind,halpernetal}. 
\par
Since its birth, Matrix theory has enjoyed numerous successes in a variety 
 of contexts, for which we refer the reader to an assortment of
 review articles \cite{reviews}. 
A simple yet fundamental question is \lq\lq why ?"
Although a number of appealing reasons were presented in the original
 proposal \cite{bfss} and  additional supportive arguments were subsequently
 supplied in \cite{seiberg, sen}, it is fair to say that the 
 precise reason is yet to be identified. 
\parsmallskip
Undoubtedly, one of the key ingredients must be the high degree of 
 SUSY present in the theory.  Besides playing the crucial role of 
 allowing the scattering states to exist, it supports a variety of
stable brane configurations expected of a proper representation 
 of M-theory. Furthermore, evidence has been accumulating \cite{Pabanetal1,
Pabanetal2,lowe,ss,Hyunetal,np}
 that at least for some restricted configurations it appears to 
 be powerful enough to control the dynamical details of the scattering. 
This is rather surprising since normally such a global symmetry 
 can only give certain relations among correlation functions and 
 cannot fix the scattering amplitudes.  
\parsmallskip
Thus it is of great interest to know precisely how much is 
 determined by SUSY. This requires investigation of the 
off-shell correlation functions as well, \ie of the effective 
 action without imposing the equations of motion. 
In contrast to the case with a small number of 
 SUSY, the theory with maximal SUSY, such as Matrix theory under 
 consideration, presents difficulties in answering to this question
 because the off-shell unconstrained superfield formalism is not known and 
 one is forced to deal with the component formalism. In such 
 non-covariant formulation, the SUSY algebra gets intimately intertwined 
 with gauge symmetry and does not close without the aid of the equations of 
 motion. Besides, the SUSY transformations for the effective action get 
 non-trivial quantum corrections, which makes the analysis quite cumbersome. 
\parsmallskip
Nevertheless, in previous works we have made some progress in understanding
 the role of SUSY for the fully off-shell effective action  for the 
 so-called the source-probe configuration, where a probe D-particle,
 with its position and the spin described respectively by the bosonic 
 coordinates $r_m(\tau)$ and the fermionic coordinates $\theta_\al(\tau)$, 
 interacts with a large number of source D-particles all sitting at
 the origin. First the precise form of the SUSY Ward identity, mixed with
 the gauge structure in essential ways, was derived together with 
  closed form expressions for the quantum-corrected SUSY transformation
 laws \cite{Kaz-Mura1}. This formalism was then explicitly 
applied to the simplest case, namely to the 1-loop effective action 
 at order 2 in the sense of the derivative expansion\footnote{The 
 concept of \lq\lq order" in the derivative expansion is defined, as 
 usual, as the number of time derivatives plus half the number of fermions.
 For the importance of the derivative expansion scheme, 
 see the detailed discussions presented in \cite{Kaz-Mura1}.}, and
 it was found that, with the knowledge of the SUSY transformation, 
 the SUSY Ward identity completely fixed the form of the off-shell 
effective action to the order investigated. This example, however, was
 not so significant since at this order the higher order derivatives,
 such as acceleration etc., that do not vanish by the equation of motion 
 nevertheless can be eliminated by integration by parts. 
\parsmallskip
The full significance of our off-shell machinery becomes apparent at order 4,
 where complete elimination of higher derivatives is no longer possible. 
Since, unlike for order 2, only a small portion  of 
  the fully off-shell 1-loop 
effective action $\Ga^{(1)}$ was known 
 at this order, we had to first  complete its 
 computation, including all the spin effects \cite{Kaz-Mura2}. This was 
 a horrendous task, involving manipulations of hundreds of terms 
in the intermediate stages. However, with judicious use of 
 $SO(9)$ Fierz identities, including some new ones, the result turned 
 out to be much simpler than was expected. To check this result as well as
 for its own interest, we also computed the quantum-corrected SUSY 
 transformation operator $\delta_\ep$ up to the relevant order. 
 Although their forms were embarrassingly complicated, 
we succeeded in verifying the Ward identity $\delta_\ep
 (\Ga^{(0)}+\Ga^{(1)})=0$, where $\Ga^{(0)}$
 is the tree-level action. 
\parsmallskip
The main purpose of the present work is to demonstrate that, conversely,
 given  such $\delta_\ep$ our $\Ga^{(1)}$ is in fact the unique 
 solution to the SUSY Ward identity at one-loop at order 4. We shall show this
 in the following way: Let us write the most general effective action
 up to 1-loop as $\Ga=\Ga^{(0)} + \Ga^{(1)} + \Delta \Ga^{(1)}$,
 where $\Ga^{(1)}$ is our previous result and $\Delta\Ga^{(1)}$ 
 is a possible arbitrary addition at order 4.
 Then, expanding $\delta_\ep$  also to 1-loop as 
$\delta_\ep =\delta_\ep^{(0)}+\delta_\ep^{(1)}$, 
the Ward identity $\delta_\ep \Ga=0$ is satisfied to the desired order 
if and only if  $\delta^{(0)}_\ep \Delta\Ga^{(1)} =0$. 
Writing  $\Delta\Ga^{(1)} =\int d\tau \Delta \calL^{(1)}$, this is 
equivalent to 
\begin{align}
 \delta^{(0)}_\ep \Delta \calL^{(1)} = \frac{dX}{d\tau} \, ,
\label{basiceq}
\end{align}
where $X$ is an 
arbitrary  expression of order 3. What we shall do is to generate 
the most general expressions for $\Delta \calL^{(1)}$ and $X$ 
allowed by $SO(9)$ and discrete CPT symmetries and demonstrate that 
the only solution is $\Delta \calL^{(1)}=0$ up to 
a total derivative, \ie $\Delta\Ga^{(1)} =0$. 

It should be emphasized that our method 
 actually  {\it does not} rely on the detailed forms of $\Ga^{(1)}$
 and $\delta_\ep^{(1)}$. We may take this as a strong indication 
that this uniqueness property  would persist to higher orders. 
Although  restricted to the source-probe configuration, 
our result reveals in a clear way the unusual power of maximal 
supersymmetry in  controlling  even the off-shell structures of Matrix theory. 
\parsmallskip
The organization of the rest of the article is as follows:
In Sec.~2, we give a brief description of Matrix theory and its
 \lq\lq source-probe" configuration, to be studied in this article. 
It includes the explanation of how supersymmetry, $SO(9)$ symmetry 
 and discrete C-P-T symmetries act in this setting. 
The analysis of the SUSY Ward identity is performed in Sec.~3. 
After elucidating our strategy, we describe how to generate 
 the most general form of $\Delta\calL^{(1)}$ and 
the possible total derivative terms in its SUSY transform and reduce 
them by the use of various 
 Fierz identities to truly independent structures. 
Using the results of these analyses,
 the SUSY Ward identity  (\ref{basiceq}) is transformed into 
 sets of linear coupled equations for the coefficients of such
 structures. Upon solving them we shall find that $\Delta \calL^{(1)}$ 
 must identically vanish, demonstrating the uniqueness of the solution 
 of the Ward identity at 1-loop at order 4. 
Sec.~4 is devoted to discussion of the implication of our result 
and some further remarks. In the Appendix, the Fierz identities 
 used in the main text are collected. 
\section{Preliminaries}
The classical action for the $U(N+1)$ Matrix theory in the Euclidean
 formulation is given by
\begin{align}
S_0  &=  \!\int\!d\tau \, \Tr 
\left(
\frac{1}{2}[D_\tau ,X^m]^2 - \frac{g^2}{4}([X^m , X^n])^2 
+ \frac{1}{2} \Theta^T [D_\tau , \Theta] 
- \frac{g}{2}\Theta^T\gamma^m [X_m,\Theta]\right)\comma \\
D_\tau  &\equiv \del_\tau -igA \period
\end{align}
In this expression, $X^m_{ij}(\tau), A_{ij}(\tau)$ and
$\Theta_{\al,ij}(\tau)$ are the $(N+1)\times (N+1)$ hermitian matrix
fields, representing the bosonic part of the D-particles, the gauge
fields, and the fermionic part of the D-particles, respectively.
$D_\tau$ is the covariant derivative, $\ga^m$ are
the real symmetric $16\times 16$ $SO(9)$ $\ga$-matrices, and the vector
index $m$ runs from $1$ to $9$.

Among the number of important symmetries possessed by this action,
 our main focus will be the supersymmetry. 
It carries 16 spinorial parameters $\ep_\al$ and transforms 
the basic fields as
\begin{align}
\delta_\epsilon A &=  \epsilon^T \Theta \comma \qquad 
\delta_\epsilon X^m = -i\epsilon^T \gamma^m\Theta \comma \label{susyb}\\
\delta_\epsilon \Theta &= i\left( \com{D_\tau}{X_m} \gamma^m
 +{g \over 2} \com{X_m}{X_n} \ga^{mn} \right) \ep\label{susyf} \, . 
\end{align}
Although the SUSY algebra closes only on-shell up to field-dependent
gauge transformations, $S_0$ {\it is} invariant under (\ref{susyb}) and
(\ref{susyf})\footnote{Strictly speaking, this symmetry is not
supersymmetry since the bose and fermi dergrees of freedom do not match
off-shell; it is a symmetry that becomes SUSY on-shell. Nevertheless, in
this article we shall call (\ref{susyb}) and (\ref{susyf}) SUSY
transformations, following common usage.}.

For the present investigation, two additional symmetries 
 play crucial roles.  One is the $SO(9)$ symmetry. Besides giving restrictions
 on the forms of various quantities, it provides highly non-trivial relations 
 among  terms involving fermions through numerous Fierz identities. 
The other is the discrete C-P-T symmetries inherited from the 10-dimensional 
super Yang-Mills theory, from which the above action can be obtained by 
dimensional reduction. To understand these symmetries properly, 
 tentatively we go back to the Minkowski formulation. 
Since the fermions in Matrix theory are Weyl-projected
 (as well as Mojorana), P and T are separately violated 
while C and CPT symmetries remain preserved. 
 Under the C-transformation, the Minkowski fields transform as
\begin{align}
\Theta_\alpha \to  \Theta_\alpha^T \, , \ \   
A_0 \to - A_0^T \, , \ \ X_m \to - X_m^T\comma 
\end{align}
where the superscript $T$ denotes the matrix-transposition. 
On the other hand, one finds that the CPT-transformation does not 
transform the fields but flips the sign of the time-derivative and 
effects $i\rightarrow -i$ as it is anti-unitary. 
\parsmallskip
Now let us give a brief description of the \lq\lq source-probe" 
 situation, on which our investigation will be focused. 
It is the configuration of a (probe) D-particle 
interacting with a large number, $N$, of  (source) D-particles all sitting 
 at the origin. This is expressed by the splitting 
\begin{align}
X^m(\tau) & = \frac{1}{g}B^m(\tau) + Y^m(\tau) \, , 
& \Theta^\alpha(\tau) & = \frac{1}{g}\theta^\alpha(\tau)
 + \Psi^\alpha(\tau) , \\
B^m(\tau) & = {\rm diag}(
r^m(\tau) ,\underbrace{0, \cdots, 0}_{\displaystyle N}) \, , 
& \theta^\alpha(\tau) & = {\rm diag}
(\theta^\alpha(\tau),\underbrace{0, \cdots, 0}_{\displaystyle N})\comma 
\end{align}
where $B_m(\tau)$ and $\theta_\al(\tau)$ are the bosonic and the
 fermionic backgrounds expressing the positions and the spin degrees of
 freedom of the D-particles respectively and $Y_m(\tau)$ and
 $\Psi_\al(\tau)$ denote the quantum fluctuations around them. We will
 be interested in the fully off-shell 
effective action $\Ga[r^m,\theta_\al]$ of the probe 
 D-particle after all the fluctuations are integrated out. In other words, 
we shall not impose any equations of motion for 
$r_m(\tau)$ and  $\theta_\al(\tau)$, 
\ie they will be {\it arbitrary functions of $\tau$}. At the tree level,
 the effective action and the SUSY transformations take the form 
\begin{align}
\widetilde{\Gamma}^{(0)} & = 
\int \! d\tau \left(
\frac{v^2}{2 g^2} + \frac{\theta\dot{\theta}}{2 g^2} \right) \, , \\
\delta_\epsilon^{(0)} r^m & = - i \epsilon \gamma^m \theta \, , \ \ \quad
\delta_\epsilon^{(0)} \theta_\alpha = i (\dirac{v}\epsilon)_\alpha \period
\end{align}
Here and hereafter, the dot means differentiation with respect to 
 the Euclidean time $\tau$ and we will use $v^m$ and $a^m$ to denote 
$\dot{r}^m$ and $\ddot{r}^m$ respectively. Contractions of the spinor indices 
 are often suppressed, so that $\theta\dot{\theta}$ stands for 
$\theta_\alpha \dot \theta_\alpha$, etc. 

For the source-probe configuration described above, the restrictions imposed 
 on the effective action by the C and  CPT symmetries, the $SO(9)$ symmetry 
and the hermiticity requirement can be worked out in the following way. 
Again to avoid misunderstanding we must first go back to the Minkowski 
formulation, in which the time $t$ is related to the Euclidean time $\tau$
 by $t=-i\tau$,  and then translate the results into the Euclidean language. 
\begin{enumerate}
 \item The C-transformation only affects $r^m$ as  $r^m \to - r^m$ 
and dictates that the number of $r^m$'s must be even.
 \item The $SO(9)$ invariance imposes two types of restrictions. 
First, the spinors must come in bilinears so that the spinor indices 
 are contracted.  Next, since 
the vector indices must also be contracted to produce  $SO(9)$ singlets,
the number of such indices must be even. This together with the 
C-invariance above dictates that the total number of $\gamma^m$'s must be even. %
\item Let $\theta^{(m)}$ denote the $m$-th $t$-derivative of $\theta$. Then, 
any fermion bilinear of the form $\theta^{(m)} A \theta^{(n)}$
 with $A$ a real matrix  (such as $\ga^{i_1i_2\ldots i_k}$) 
 is anti-hermitian since 
$(\theta^{(m)} A\theta^{(n)})^\dagger
 = \theta^{(n)} A^T\theta^{(m)} = -\theta^{(m)} A\theta^{(n)}$.
Thus, a hermitian spinor bilinear, to be used as a building block, 
 must be of the 
form $i \theta^{(m)} A \theta^{(n)}$, \ie with an overall $i$. 
 \item  Now consider the CPT transformation. 
On the bosonic coordinates, its only effect is to change the sign of the time 
derivative and hence we have 
$r^m \rightarrow r^m$, $v^m \to - v^m$, etc. This is often referred to as 
\lq\lq time-reversal" from the M-theory point of view. 
It gives the  well known restriction that the number of time-derivatives 
in the purely bosonic terms must be even. On the other hand, due to an 
 overall $i$, CPT transform of a {\it hermitian} fermion bilinear 
$i \theta^{(m)} A \theta^{(n)}$ acquires an extra minus sign from 
 $i \rightarrow -i$, apart from the factors of $-1$ from the
 time-derivatives. This means that we can count  $\theta^2$
 as one time-derivative as far as  the CPT invariance is concerned.
 Combined with the requirement 
 for the bosonic part, this amounts to a remarkably simple statement 
 that CPT invariance is equivalent to demanding  the ``order'', as 
defined in the sense of the derivative expansion, to be even. 
\end{enumerate}
Going back to the Euclidean formulation by $t\rightarrow -i\tau$, 
 we can summarize the above requirements as follows:
\begin{itemize}
	\item In constructing the effective Lagrangian,
 use $i^{1+m+n}\theta^{(m)} \ga^{i_1i_2\ldots i_k}\theta^{(n)}$
 as fermionic building block and demand that the number of $r^m$, the number
 of $\ga^m$ and the \lq\lq order" be all even. 
\end{itemize}
This rule will facilitate the generation of the most general effective 
 action,  to be performed in the next section. 
\section{Analysis of the SUSY Ward identity}
\subsection{Strategy}
We begin the analysis of the SUSY Ward identity by explaining our
 strategy. As was already explained in the introduction,
in terms of the unintegrated quantities the Ward identity we wish to 
 analyze is 
\begin{align}
\delta_\epsilon^{(0)}\Delta \calL^{(1)} = \frac{d X}{d \tau}\comma
\label{treeinv}
\end{align}
where $\Delta\calL^{(1)}$ is the most general 
effective Lagrangian possible at 1-loop at order 4 and 
 $X$ stands for an arbitrary expression at order 3. 
To solve this equation, we expand 
$\Delta\calL^{(1)}$ and $X$ in powers  of $\theta$ in the manner
\begin{align}
 \Delta\calL^{(1)} &\equiv 
 \Delta\calL^{(1)}_{\theta^0} +
 \Delta\calL^{(1)}_{\theta^2} +
 \Delta\calL^{(1)}_{\theta^4} +
 \Delta\calL^{(1)}_{\theta^6} +
 \Delta\calL^{(1)}_{\theta^8} ,\label{expl}\\
X &\equiv  
X_{\epsilon\p^3\theta} + 
X_{\epsilon\p^2\theta^3} + 
X_{\epsilon\p\theta^5} + 
X_{\epsilon\theta^7}\period \label{expx}
\end{align}
The subscripts indicate the structure of each term schematically and 
 should be self-explanatory. Now the tree-level SUSY variation
 $\delta^{(0)}_\ep$ consists of  $\delta^{(0)}_r$ which 
 acts on $r^m$ to add one $\theta$ and $\delta^{(0)}_\theta$
 that acts on $\theta_\al$ to remove one $\theta$. Thus, by 
substituting the expansions (\ref{expl}) and (\ref{expx}) into the equation
(\ref{treeinv}) and collecting terms with the same number of $\theta$'s, 
the Ward identity can be split into the following five equations: 
\begin{align}
& \delta_r^{(0)} \Delta\calL^{(1)}_{\theta^0}
+
\delta_\theta^{(0)} \Delta\calL^{(1)}_{\theta^2} = 
\frac{d X_{\epsilon\p^3\theta}}{d \tau}, 
\label{treesusy0}
  \\
& \delta_r^{(0)} \Delta\calL^{(1)}_{\theta^2}
+
\delta_\theta^{(0)} \Delta\calL^{(1)}_{\theta^4} = 
\frac{d X_{\epsilon\p^2\theta^3}}{d \tau},  
\label{treesusy2}\\
& \delta_r^{(0)} \Delta\calL^{(1)}_{\theta^4}
+
\delta_\theta^{(0)} \Delta\calL^{(1)}_{\theta^6} = 
\frac{d X_{\epsilon\p\theta^5}}{d \tau},  
\label{treesusy4}\\
& \delta_r^{(0)} \Delta\calL^{(1)}_{\theta^6}
+
\delta_\theta^{(0)} \Delta\calL^{(1)}_{\theta^8} =
\frac{d X_{\epsilon\theta^7}}{d \tau}, 
\label{treesusy6} \\
& \delta_r^{(0)} \Delta\calL^{(1)}_{\theta^8} = 0\period
\label{treesusy8} 
\end{align}
Our task then is to solve each of these equations to determine 
 $\Delta \calL^{(1)}_{\theta^n}$. Since {\it both} $\Delta \calL_{\theta^n}$'s 
 {\it and} $X_{\ep\del^n\theta^m}$'s are unknown, except for restrictions 
 imposed by $SO(9)$ and CPT symmetries, one might at first sight suspect
 that the solution would not be unique. As we shall see,  this expectation 
 will be refuted. 
\subsection{General form of the effective Lagrangian}
We will generate the most general expressions for 
$\Delta \calL^{(1)}_{\theta^n}$'s in following steps: 
First, we  enumerate all possible structures contributing 
 to $\Delta \calL^{(1)}_{\theta^n}$ at order 4, taking into account the 
 symmetry requirement spelled out at the end of the previous section. 
Next we will eliminate all possible total derivative combinations. 
Finally, for terms containing 4 $\theta$'s or more, we apply 
appropriate $SO(9)$ Fierz identities to reduce the expressions to 
 truly independent ones. 

Let us describe this  procedure for each $\Delta \calL^{(1)}_{\theta^n}$ 
more explicitly. In the following, $A_i \sim E_i$ will denote
 arbitrary constants. 
\parmedskipn
{\bf Expression for  $\Delta \calL^{(1)}_{\theta^0}$}:\parmedskip
For the purely bosonic part, it is straightforward to see that 
 there are  6 independent terms, which can be chosen to be 
\begin{align}
\Delta\calL^{(1)}_{\theta^0} =
 \frac{{A_1}\,{(r \cdot v)}^4}{r^{11}}
   +\frac{{A_2}\,{(r \cdot a)}^2}{r^7}
   +\frac{{A_3}\,v^2\,{(r \cdot v)}^2}{r^9}
   +\frac{{A_4}\,v^2\,(r \cdot a)}{r^7}
   +\frac{{A_5}\,v^4}{r^7}
   +\frac{{A_6}\,a^2}{r^5} \period
\label{eff44}
\end{align}
\parn
{\bf Expression for  $\Delta \calL^{(1)}_{\theta^2}$}:\parmedskip
Similarly, it is not difficult to enumerate all possible terms 
 with two $\theta$'s and we find the following 11 independent terms:
\begin{align}
\Delta\calL^{(1)}_{\theta^2} = 
& 
+ \frac{{B_1}\,({\ddot{\theta}}{\dot{\theta}})}{r^5}
   +\frac{{B_2}\,{(r \cdot v)}^2\,({\dot{\theta}}\theta )}{r^9}
   +\frac{{B_3}\,(r \cdot a)\,({\dot{\theta}}\theta )}{r^7}
   +\frac{{B_4}\,v^2\,({\dot{\theta}}\theta )}{r^7}
   +\frac{{B_5}\,{r_j}\,{v_i}\,
       ({\dot{\theta}}{{\gamma }^{ij}}{\dot{\theta}})}{r^7}
\brkeq 
+
    \frac{{B_6}\,{r_j}\,{v_i}\,(r \cdot v)\,
       ({\dot{\theta}}{{\gamma }^{ij}}\theta )}{r^9}+
    \frac{{B_7}\,{(r \cdot v)}^2\,{r_j}\,{v_i}\,
       (\theta {{\gamma }^{ij}}\theta )}{r^{11}} + 
    \frac{{B_8}\,v^2\,{r_j}\,{v_i}\,
       (\theta {{\gamma }^{ij}}\theta )}{r^9} \brkeq 
+
    \frac{{B_9}\,{r_j}\,{a_i}\,
       ({\dot{\theta}}{{\gamma }^{ij}}\theta )}{r^7} 
+
    \frac{{B_{10}}\,{r_j}\,{a_i}\,(r \cdot v)\,
       (\theta {{\gamma }^{ij}}\theta )}{r^9}
   +\frac{{B_{11}}\,{v_i}\,{a_j}\,
       (\theta {{\gamma }^{ij}}\theta )}{r^7} \period
\label{eff43}
\end{align}
\parn
{\bf Expression for  $\Delta \calL^{(1)}_{\theta^4}$}:\parmedskip
Beginning at this order with 4 $\theta$'s, 
 our task becomes much more difficult. 
In addition to the number of allowed terms getting large, 
 we must find judicious Fierz identities to 
 reduce them to independent expressions. Since these identities 
 are rather involved, we shall relegate them to Appendix. 
They are generated  by an efficient new algorithm developed in the 
previous work\cite{Kaz-Mura2}. 

After eliminating terms related by total derivatives, the ones 
 contributing to $\Delta \calL^{(1)}_{\theta^4}$ are found to be 
\begin{align}
 \and \frac{(\theta {{\gamma }^{{a_1}{a_2}}}\theta )\,
      ({\dot{\theta}}{{\gamma }^{{a_1}{a_2}}}{\dot{\theta}})
      }{r^7},\frac{(\theta {{\gamma }^{{a_1}{a_2}{a_3}}}
       \theta )\,({\dot{\theta}}
       {{\gamma }^{{a_1}{a_2}{a_3}}}{\dot{\theta}})}{r^7},
   \frac{(r \cdot v)\,(\theta {{\gamma }^{{a_1}{a_2}}}
       \theta )\,({\dot{\theta}}{{\gamma }^{{a_1}{a_2}}}
       \theta )}{r^9}, \brkeq
   \frac{(r \cdot v)\,(\theta {{\gamma }^{{a_1}{a_2}{a_3}}}
       \theta )\,({\dot{\theta}}
       {{\gamma }^{{a_1}{a_2}{a_3}}}\theta )}{r^9},
   \frac{{(r \cdot v)}^2\,
      {(\theta {{\gamma }^{{a_1}{a_2}}}\theta )}^2}{r^{11}},
   \frac{{(r \cdot v)}^2\,
      {(\theta {{\gamma }^{{a_1}{a_2}{a_3}}}\theta )}^2}
      {r^{11}}, \brkeq
   \frac{{r_{{i_1}}}\,{r_{{i_2}}}\,{(r \cdot v)}^2\,
      (\theta {{\gamma }^{{a_1}{i_1}}}\theta )\,
      (\theta {{\gamma }^{{a_1}{i_2}}}\theta )}{r^{13}},
   \frac{{r_{{i_1}}}\,{r_{{i_2}}}\,
      (\theta {{\gamma }^{{a_1}{i_2}}}\theta )\,
      ({\dot{\theta}}{{\gamma }^{{a_1}{i_1}}}{\dot{\theta}})
      }{r^9},\frac{{r_{{i_1}}}\,{r_{{i_2}}}\,(r \cdot v)\,
      (\theta {{\gamma }^{{a_1}{i_2}}}\theta )\,
      ({\dot{\theta}}{{\gamma }^{{a_1}{i_1}}}\theta )}{r^
      {11}}, \brkeq\frac{{r_{{i_1}}}\,{r_{{i_2}}}\,
      {(r \cdot v)}^2\,(\theta {{\gamma }^{{a_1}{a_2}{i_1}}}
       \theta )\,(\theta {{\gamma }^{{a_1}{a_2}{i_2}}}
       \theta )}{r^{13}},
   \frac{{r_{{i_1}}}\,{r_{{i_2}}}\,
      (\theta {{\gamma }^{{a_1}{a_2}{i_2}}}\theta )\,
      ({\dot{\theta}}{{\gamma }^{{a_1}{a_2}{i_1}}}
       {\dot{\theta}})}{r^9},    \frac{v^2\,{(\theta {{\gamma }^{{a_1}{a_2}}}\theta )}^2}
    {r^9},
\brkeq
\frac{v^2\,{(\theta {{\gamma }^{{a_1}{a_2}{a_3}}}
         \theta )}^2}{r^9},
   \frac{{r_{{i_1}}}\,{r_{{i_2}}}\,(r \cdot v)\,
      (\theta {{\gamma }^{{a_1}{a_2}{i_2}}}\theta )\,
      ({\dot{\theta}}{{\gamma }^{{a_1}{a_2}{i_1}}}\theta )}
      {r^{11}}, 
   \frac{v^2\,{r_{{i_1}}}\,{r_{{i_2}}}\,
      (\theta {{\gamma }^{{a_1}{i_1}}}\theta )\,
      (\theta {{\gamma }^{{a_1}{i_2}}}\theta )}{r^{11}},
    \brkeq\frac{v^2\,{r_{{i_1}}}\,{r_{{i_2}}}\,
      (\theta {{\gamma }^{{a_1}{a_2}{i_1}}}\theta )\,
      (\theta {{\gamma }^{{a_1}{a_2}{i_2}}}\theta )}{r^{11}}
    ,\frac{{r_{{i_1}}}\,{v_{{i_2}}}\,
      ({\dot{\theta}}\theta )\,
      (\theta {{\gamma }^{{i_1}{i_2}}}\theta )}{r^9},
   \frac{{r_{{i_1}}}\,{v_{{i_2}}}\,(r \cdot v)\,
      (\theta {{\gamma }^{{a_1}{i_1}}}\theta )\,
      (\theta {{\gamma }^{{a_1}{i_2}}}\theta )}{r^{11}},
    \brkeq\frac{{r_{{i_1}}}\,{v_{{i_2}}}\,(r \cdot v)\,
      (\theta {{\gamma }^{{a_1}{a_2}{i_1}}}\theta )\,
      (\theta {{\gamma }^{{a_1}{a_2}{i_2}}}\theta )}{r^{11}}
    ,\frac{{r_{{i_1}}}\,{v_{{i_2}}}\,
      (\theta {{\gamma }^{{a_1}{i_1}{i_2}}}\theta )\,
      ({\dot{\theta}}{{\gamma }^{{a_1}}}\theta )}{r^9},
   \frac{{r_{{i_1}}}\,{v_{{i_2}}}\,
      (\theta {{\gamma }^{{a_1}{i_1}}}\theta )\,
      ({\dot{\theta}}{{\gamma }^{{a_1}{i_2}}}\theta )}{r^9},
    \brkeq\frac{{r_{{i_1}}}\,{v_{{i_2}}}\,
      (\theta {{\gamma }^{{a_1}{i_2}}}\theta )\,
      ({\dot{\theta}}{{\gamma }^{{a_1}{i_1}}}\theta )}{r^9},
   \frac{{r_{{i_1}}}\,{v_{{i_2}}}\,
      (\theta {{\gamma }^{{a_1}{a_2}{i_1}}}\theta )\,
      ({\dot{\theta}}{{\gamma }^{{a_1}{a_2}{i_2}}}\theta )}
      {r^9},\frac{{r_{{i_1}}}\,{v_{{i_2}}}\,
      (\theta {{\gamma }^{{a_1}{a_2}}}\theta )\,
      ({\dot{\theta}}{{\gamma }^{{a_1}{a_2}{i_1}{i_2}}}
       \theta )}{r^9}, \brkeq
   \frac{{v_{{i_1}}}\,{v_{{i_2}}}\,
      (\theta {{\gamma }^{{a_1}{i_1}}}\theta )\,
      (\theta {{\gamma }^{{a_1}{i_2}}}\theta )}{r^9},
   \frac{{r_{{i_1}}}\,{v_{{i_2}}}\,
      (\theta {{\gamma }^{{a_1}{a_2}{i_2}}}\theta )\,
      ({\dot{\theta}}{{\gamma }^{{a_1}{a_2}{i_1}}}\theta )}
      {r^9},\frac{{v_{{i_1}}}\,{v_{{i_2}}}\,
      (\theta {{\gamma }^{{a_1}{a_2}{i_1}}}\theta )\,
      (\theta {{\gamma }^{{a_1}{a_2}{i_2}}}\theta )}{r^9},
    \brkeq
\frac{{r_{{i_1}}}\,{r_{{i_2}}}\,{v_{{i_3}}}\,
      {v_{{i_4}}}\,(\theta {{\gamma }^{{a_1}{i_1}{i_3}}}
       \theta )\,(\theta {{\gamma }^{{a_1}{i_2}{i_4}}}
       \theta )}{r^{11}},
\frac{{r_{{i_1}}}\,{r_{{i_2}}}\,{v_{{i_3}}}\,
      {v_{{i_4}}}\,(\theta {{\gamma }^{{i_1}{i_3}}}
       \theta )\,(\theta {{\gamma }^{{i_2}{i_4}}}
       \theta )}{r^{11}}, \brkeq
 \frac{{r_{{i_1}}}\,{v_{{i_2}}}\,
      (\theta {{\gamma }^{{a_1}{a_2}{a_3}}}\theta )\,
      ({\dot{\theta}}{{\gamma }^{{a_4}{a_5}{a_6}{a_7}}}
       \theta )\,{{\epsilon }_
        {{a_1}{a_2}{a_3}{a_4}{a_5}{a_6}{a_7}{i_1}{i_2}}}}
      {r^9}\period
\end{align}
There are altogether 30 terms, which are not yet all independent. 
We now indicate how they can be reduced to the independent ones 
by the use of Fierz identities. 

First, several terms in the above list  simply vanish 
 due to the Fierz identities (\ref{F01}) and (\ref{F02}). 
Next, the last term, the only term with the $\ep$-tensor, can be
 rewritten into an expression  without it by using the identity 
(\ref{F2ep1}). It becomes 
\begin{align}
 \frac{{r_{{i_1}}}\,{v_{{i_2}}}\,
    (\theta {{\gamma }^{{a_1}{a_2}{a_3}}}\theta )\,
    ({\dot{\theta}}{{\gamma }^{{a_4}{a_5}{a_6}{a_7}}}\theta 
     )\,{{\epsilon }_
      {{a_1}{a_2}{a_3}{a_4}{a_5}{a_6}{a_7}{i_1}{i_2}}}}{r^9}
= 720 \frac{{r_{{i_1}}}\,{v_{{i_2}}}\,({\dot{\theta}}\theta )\,
    (\theta {{\gamma }^{{i_1}{i_2}}}\theta )}{r^9} .
\end{align}
Similarly, we can relate various other terms by using the identities
 (\ref{F03}) $\sim$ (\ref{F41}).

After all possible such manipulations, the number of independent 
 structures is reduced to 13 and we get 
\begin{align}
 \Delta\calL^{(1)}_{\theta^4} =
 \and \frac{{C_1}\,{({\dot{\theta}}\theta )}^2}{r^7}
   +\frac{{C_2}\,{({\dot{\theta}}{{\gamma }^{i}}\theta )}^2}
     {r^7}
   +\frac{{C_3}\,{r_i}\,{r_j}\,
       ({\dot{\theta}}{{\gamma }^{i}}\theta )\,
       ({\dot{\theta}}{{\gamma }^{j}}\theta )}{r^9}
+
    \frac{{C_4}\,{r_j}\,{r_k}\,
       (\theta {{\gamma }^{ij}}\theta )\,
       ({\dot{\theta}}{{\gamma }^{ik}}{\dot{\theta}})}{r^9}
\brkeq 
   +\frac{{C_5}\,{r_j}\,{r_k}\,(r \cdot v)\,
       (\theta {{\gamma }^{ij}}\theta )\,
       ({\dot{\theta}}{{\gamma }^{ik}}\theta )}{r^{11}}
+     \frac{{C_6}\,{r_j}\,{r_k}\,{(r \cdot v)}^2\,
       (\theta {{\gamma }^{ij}}\theta )\,
       (\theta {{\gamma }^{ik}}\theta )}{r^{13}} \brkeq 
+
    \frac{{C_7}\,{r_j}\,{r_k}\,v^2\,
       (\theta {{\gamma }^{ij}}\theta )\,
       (\theta {{\gamma }^{ik}}\theta )}{r^{11}}
+\frac{{C_8}\,{r_i}\,{v_j}\,
       ({\dot{\theta}}\theta )\,
       (\theta {{\gamma }^{ij}}\theta )}{r^9} 
+
    \frac{{C_9}\,{r_j}\,{v_k}\,
       (\theta {{\gamma }^{ik}}\theta )\,
       ({\dot{\theta}}{{\gamma }^{ij}}\theta )}{r^9} \brkeq 
   +\frac{{C_{10}}\,{r_j}\,{v_k}\,
       (\theta {{\gamma }^{ij}}\theta )\,
       ({\dot{\theta}}{{\gamma }^{ik}}\theta )}{r^9}
+
    \frac{{C_{11}}\,{r_j}\,{v_k}\,(r \cdot v)\,
       (\theta {{\gamma }^{ij}}\theta )\,
       (\theta {{\gamma }^{ik}}\theta )}{r^{11}} \brkeq 
+
    \frac{{C_{12}}\,{v_j}\,{v_k}\,
       (\theta {{\gamma }^{ij}}\theta )\,
       (\theta {{\gamma }^{ik}}\theta )}{r^9}
   +\frac{{C_{13}}\,{r_i}\,{r_k}\,{v_j}\,{v_l}\,
       (\theta {{\gamma }^{ij}}\theta )\,
       (\theta {{\gamma }^{kl}}\theta )}{r^{11}}\period
\label{eff42}
\end{align}
{\bf Expression for  $\Delta \calL^{(1)}_{\theta^6}$}:\parmedskip
Now we come to the structures with 6 $\theta$'s. The number of possible
 terms, disregarding total derivatives, turned out to be 45. 
As they are too space-filling to be displayed here, we shall only 
sketch the reduction steps. 

First, a number of terms containing an $\ep$-tensor can all be 
turned into those without one by the identities (\ref{F3ep2})
 $\sim$ (\ref{F3ep1}) and (\ref{F2ep1}).  The remaining terms can then be 
 reduced  by using the identities (\ref{F00}) $\sim$ (\ref{F42}) 
 as well as those already used at $\calO(\theta^4)$. 

These manipulations drastically reduce the number of independent 
 terms and we end up with the following simple expression: 
\begin{align}
\Delta\calL^{(1)}_{\theta^6} = 
\and \frac{{D_1}\,{r_j}\,{r_k}\,({\dot{\theta}}\theta )\,
       (\theta {{\gamma }^{ij}}\theta )\,
       (\theta {{\gamma }^{ik}}\theta )}{r^{11}}+
    \frac{{D_2}\,{r_k}\,{r_l}\,
       (\theta {{\gamma }^{ij}}\theta )\,
       (\theta {{\gamma }^{ik}}\theta )\,
       ({\dot{\theta}}{{\gamma }^{jl}}\theta )}{r^{11}}
    \brkeq +\frac{{D_3}\,{r_l}\,{v_k}\,
       (\theta {{\gamma }^{ij}}\theta )\,
       (\theta {{\gamma }^{ik}}\theta )\,
       (\theta {{\gamma }^{jl}}\theta )}{r^{11}}+
    \frac{{D_4}\,{r_j}\,{r_k}\,{r_l}\,{v_m}\,
       (\theta {{\gamma }^{ij}}\theta )\,
       (\theta {{\gamma }^{ik}}\theta )\,
       (\theta {{\gamma }^{lm}}\theta )}{r^{13}}   \period
\label{eff41}
\end{align}
{\bf Expression for  $\Delta \calL^{(1)}_{\theta^8}$}:\parmedskip
Finally, we are left with the expression with 8 $\theta$'s, the maximum
 number allowed at order 4. 
Starting from 36 possible such 
 expressions, application of appropriate Fierz identities 
listed in the Appendix leads to 
\begin{align}
\Delta\calL^{(1)}_{\theta^8} = 
\and 
 \frac{{E_1}\,(\theta {{\gamma }^{ij}}\theta )\,
       (\theta {{\gamma }^{ik}}\theta )\,
       (\theta {{\gamma }^{jl}}\theta )\,
       (\theta {{\gamma }^{kl}}\theta )}{r^{11}}
   +\frac{{E_2}\,{r_k}\,{r_m}\,
       (\theta {{\gamma }^{ij}}\theta )\,
       (\theta {{\gamma }^{ik}}\theta )\,
       (\theta {{\gamma }^{jl}}\theta )\,
       (\theta {{\gamma }^{lm}}\theta )}{r^{13}} \brkeq 
   +\frac{{E_3}\,{r_j}\,{r_k}\,{r_m}\,{r_n}\,
       (\theta {{\gamma }^{ij}}\theta )\,
       (\theta {{\gamma }^{ik}}\theta )\,
       (\theta {{\gamma }^{lm}}\theta )\,
       (\theta {{\gamma }^{ln}}\theta )}{r^{15}} \period
\label{eff40}
\end{align}
This agrees with the result of \cite{Pabanetal1, Hyunetal}, where the 
 effective action without time-derivatives of $\theta$ was considered. 

This completes the construction of the most general effective Lagrangian. 
We now turn to the generation of allowed expression for the quantity 
$X$ at order 3,  from which we produce the total derivative term $dX/d\tau$
 in the basic  Ward identity (\ref{treeinv}). 
\subsection{Generation of total derivative terms}
The basic steps for generating the general order 3 expressions $X_{\ep\del^m
\theta^n}$ are quite similar to the ones for $\Delta\calL^{(1)}_{\theta^n}$ 
described in the previous section. Fortunately, we can argue that
 of the four such $X_{\ep\del^m\theta^n}$'s  two of them, namely 
$X_{\epsilon\p^3\theta}$ and $ X_{\epsilon\p\theta^5}$, can be 
 set to zero. 
 
First, consider the terms of $\calO(\ep\del^4\theta)$ generated on the 
 left-hand side (LHS) of (\ref{treesusy0}). Schematically, they can be grouped 
 into the following 5 types:
\begin{align} 
&  Y_1[r] (\epsilon \gamma^* {\theta}^{(4)}), \ \  
  Y_2[r,v] (\epsilon \gamma^* {\theta}^{(3)}),    \ \ 
  Y_3[r,v,a](\epsilon \gamma^* \ddot{\theta}), \nn \\
&   Y_4[r,v,a,\dot{a}](\epsilon \gamma^* \dot{\theta}), \ \  
  Y_5[r,v,a,\dot{a},\ddot{a}](\epsilon \gamma^* \theta),  
\end{align} 
where $Y_1 \sim Y_5$ are functions of $r(\tau)$ and its derivatives and 
$*$ in  $\gamma^*$ stands for various sets of vector indices. 
Since total derivatives can be absorbed into $d X_{\epsilon\p^3\theta}/d\tau$
on the right-hand side (RHS), the first 4 structures can be transformed 
 into the last one in the list 
by \lq\lq integration by parts". This process is unambiguous
 since there is only one $\theta$ involved and we may simply transfer 
 the derivatives on it to bosonic variables. 
Thus, Eq. (\ref{treesusy0}) is simplified to 
\begin{align}
  Y_5[r,v,a,\dot{a},\ddot{a}](\epsilon \gamma^* \theta) = 
\frac{d X_{\epsilon\p^3\theta}}{d \tau}.
\end{align}
Now note that while the LHS has no derivatives of $\theta$, $d/d\tau$ 
 on the RHS necessarily produces $\dot{\theta}$ for each term composing 
$X_{\epsilon\p^3\theta}$. This is a contradiction unless 
$X_{\epsilon\p^3\theta}$ vanishes. 

Similar argument applies to $\calO(\ep\del^2\theta^5)$ terms in 
 the Ward identity (\ref{treesusy4}).  In this case, the structure
 of $X_{\epsilon\p\theta^5}$ is schematically
\begin{align}
X_{\epsilon\p\theta^5} \sim  Z_1[r]\epsilon \theta^4 \dot{\theta} + 
 Z_2[r]\epsilon \theta^5 \dot{r},  
\end{align}
where $Z_1[r]$ and $Z_2[r]$ are some functions of $r(\tau)$.  The time 
 derivative of this expression produces either $\ddot r^n$ or 
$\ddot \theta^\alpha$. However, the LHS of (\ref{treesusy4}) 
 can be transformed by integration by parts to expressions without 
 such double derivatives. Hence we may set $X_{\epsilon\p\theta^5} = 0$.

For the remaining two entities $X_{\ep\del^2\theta^3}$ and 
$X_{\ep\theta^7}$, no such arguments apply and we must generate all 
 possible terms which are algebraically independent. 
\pagebreak[2]
\shead{Construction of $X_{\ep\del^2\theta^3}$:}\quad 
There are 8 possible types of structures, up to total derivatives:
\begin{align}
& \text{type 1}: (\epsilon \gamma^* \theta) (\theta \gamma^* \theta)
\ddot{r}^n , \nn\\
& \text{type 2}: (\epsilon \gamma^* \theta) (\theta \gamma^* \theta)\dot{r}^n\dot{r}^m , \ \ \nn \\
& \text{type 3}: (\epsilon \gamma^* \dot{\theta}) (\theta \gamma^* \theta) \dot{r}^n , \qquad 
\text{type 3$^\prime$}: (\epsilon \gamma^*  \theta) 
(\theta \gamma^* \dot{\theta}) \dot{r}^n , \nn \\
& \text{type 4}: (\epsilon \gamma^* \ddot{\theta}) (\theta \gamma^* \theta) r^n , \qquad 
\text{type 4$^\prime$}: (\epsilon \gamma^* \theta) (\ddot{\theta} \gamma^*
\theta) r^n , \nn \\
& \text{type 5}: (\epsilon \gamma^* \theta) (\dot{\theta} \gamma^* \dot{\theta}) r^n , \qquad 
\text{type 5$^\prime$}: (\epsilon \gamma^* \dot{\theta}) (\theta \gamma^* \dot{\theta}) r^n . \nn
\end{align}
Structures of type 3$^\prime$ can be expressed in terms of those of type 3
by the use of the Fierz identities of the type 
$(\epsilon \gamma^* \theta) (\theta
\gamma^* \dot{\theta}) \to (\epsilon \gamma^* \dot{\theta}) (\theta
\gamma^* \theta) $. Similarly, type 4$^\prime$ and type 5$^\prime$
 terms can be rewritten  in terms of type 4 and type 5 terms respectively. 
Thus we only need to consider the structures of type 1 $\sim$ type 5. 

Possible type 1 terms are given by 
\begin{align}
 \and \frac{{r_{{i_1}}}\,(r \cdot a)\,
      (\epsilon {{\gamma }^{{a_1}{a_2}{i_1}}}\theta )\,
      (\theta {{\gamma }^{{a_1}{a_2}}}\theta )}{r^9},
   \frac{{a_{{i_1}}}\,(\epsilon 
       {{\gamma }^{{a_1}{a_2}{i_1}}}\theta )\,
      (\theta {{\gamma }^{{a_1}{a_2}}}\theta )}{r^7},
   \frac{{r_{{i_1}}}\,(r \cdot a)\,
      (\epsilon {{\gamma }^{{a_1}}}\theta )\,
      (\theta {{\gamma }^{{a_1}{i_1}}}\theta )}{r^9},
    \brkeq \frac{{r_{{i_1}}}\,{r_{{i_2}}}\,{a_{{i_3}}}\,
      (\epsilon {{\gamma }^{{a_1}{i_1}{i_3}}}\theta )\,
      (\theta {{\gamma }^{{a_1}{i_2}}}\theta )}{r^9},
   \frac{{a_{{i_1}}}\,(\epsilon {{\gamma }^{{a_1}}}\theta 
       )\,(\theta {{\gamma }^{{a_1}{i_1}}}\theta )}{r^7},
   \frac{{r_{{i_1}}}\,{r_{{i_2}}}\,{a_{{i_3}}}\,
      (\epsilon {{\gamma }^{{i_1}}}\theta )\,
      (\theta {{\gamma }^{{i_2}{i_3}}}\theta )}{r^9},
    \brkeq \frac{{r_{{i_1}}}\,(r \cdot a)\,
      (\epsilon {{\gamma }^{{a_1}{a_2}{a_3}{i_1}}}\theta )\,
      (\theta {{\gamma }^{{a_1}{a_2}{a_3}}}\theta )}{r^9},
   \frac{{a_{{i_1}}}\,(\epsilon 
       {{\gamma }^{{a_1}{a_2}{a_3}{i_1}}}\theta )\,
      (\theta {{\gamma }^{{a_1}{a_2}{a_3}}}\theta )}{r^7},
   \frac{{r_{{i_1}}}\,(r \cdot a)\,
      (\epsilon {{\gamma }^{{a_1}{a_2}}}\theta )\,
      (\theta {{\gamma }^{{a_1}{a_2}{i_1}}}\theta )}{r^9},
    \brkeq \frac{{r_{{i_1}}}\,{r_{{i_2}}}\,{a_{{i_3}}}\,
      (\epsilon {{\gamma }^{{a_1}{a_2}{i_1}{i_3}}}\theta )\,
      (\theta {{\gamma }^{{a_1}{a_2}{i_2}}}\theta )}{r^9},
   \frac{{a_{{i_1}}}\,(\epsilon {{\gamma }^{{a_1}{a_2}}}
       \theta )\,(\theta {{\gamma }^{{a_1}{a_2}{i_1}}}
       \theta )}{r^7},\frac{{r_{{i_1}}}\,{r_{{i_2}}}\,
      {a_{{i_3}}}\,(\epsilon {{\gamma }^{{a_1}{i_1}}}\theta 
       )\,(\theta {{\gamma }^{{a_1}{i_2}{i_3}}}\theta )}{r^
      9}\period
\end{align}
By applying the  Fierz identities (\ref{F12}) $\sim$ 
 (\ref{F34}), they can be reduced to 5 independent structures
 of the form
\begin{align}
  \and \frac{{r_{{i_1}}}\,(r \cdot a)\,
      (\epsilon {{\gamma }^{{a_1}}}\theta )\,
      (\theta {{\gamma }^{{a_1}{i_1}}}\theta )}{r^9},
   \frac{{a_{{i_1}}}\,(\epsilon {{\gamma }^{{a_1}}}\theta 
       )\,(\theta {{\gamma }^{{a_1}{i_1}}}\theta )}{r^7},
   \frac{{r_{{i_1}}}\,{r_{{i_2}}}\,{a_{{i_3}}}\,
      (\epsilon {{\gamma }^{{a_1}{i_1}{i_3}}}\theta )\,
      (\theta {{\gamma }^{{a_1}{i_2}}}\theta )}{r^9},     \brkeq 
   \frac{{r_{{i_1}}}\,{r_{{i_2}}}\,{a_{{i_3}}}\,
      (\epsilon {{\gamma }^{{i_1}}}\theta )\,
      (\theta {{\gamma }^{{i_2}{i_3}}}\theta )}{r^9},
\frac{{r_{{i_1}}}\,{r_{{i_2}}}\,{a_{{i_3}}}\,
      (\epsilon {{\gamma }^{{a_1}{i_1}}}\theta )\,
      (\theta {{\gamma }^{{a_1}{i_2}{i_3}}}\theta )}{r^9} .
\end{align}

In a similar fashion, we can write down all possible independent terms of 
type 2 $\sim$ type 5.  The details are omitted. 

Assembling all five types of structures, we obtain $X_{\epsilon\p^2\theta^3}$
consisting of  31 independent terms with unknown coefficients $F_i$:
\begin{align}
& X_{\epsilon\p^2\theta^3} =
\frac{{F_1}\,{r_{{i_1}}}\,(r \cdot a)\,
       (\epsilon {{\gamma }^{{a_1}}}\theta )\,
       (\theta {{\gamma }^{{a_1}{i_1}}}\theta )}{r^9}+
    \frac{{F_2}\,{a_{{i_1}}}\,
       (\epsilon {{\gamma }^{{a_1}}}\theta )\,
       (\theta {{\gamma }^{{a_1}{i_1}}}\theta )}{r^7}+
    \frac{{F_3}\,{r_{{i_1}}}\,{r_{{i_2}}}\,{a_{{i_3}}}\,
       (\epsilon {{\gamma }^{{a_1}{i_1}{i_3}}}\theta )\,
       (\theta {{\gamma }^{{a_1}{i_2}}}\theta )}{r^9} \brkeq
    +\frac{{F_4}\,{r_{{i_1}}}\,{r_{{i_2}}}\,{a_{{i_3}}}\,
       (\epsilon {{\gamma }^{{i_1}}}\theta )\,
       (\theta {{\gamma }^{{i_2}{i_3}}}\theta )}{r^9}+
    \frac{{F_5}\,{r_{{i_1}}}\,{(r \cdot v)}^2\,
       (\epsilon {{\gamma }^{{a_1}}}\theta )\,
       (\theta {{\gamma }^{{a_1}{i_1}}}\theta )}{r^{11}}+
    \frac{v^2\,{F_6}\,{r_{{i_1}}}\,
       (\epsilon {{\gamma }^{{a_1}}}\theta )\,
       (\theta {{\gamma }^{{a_1}{i_1}}}\theta )}{r^9} \brkeq
    +\frac{{F_7}\,{v_{{i_1}}}\,(r \cdot v)\,
       (\epsilon {{\gamma }^{{a_1}}}\theta )\,
       (\theta {{\gamma }^{{a_1}{i_1}}}\theta )}{r^9}+
    \frac{{F_8}\,{r_{{i_1}}}\,{r_{{i_2}}}\,{v_{{i_3}}}\,
       (r \cdot v)\,(\epsilon {{\gamma }^{{a_1}{i_1}{i_3}}}
        \theta )\,(\theta {{\gamma }^{{a_1}{i_2}}}\theta )}
       {r^{11}} \brkeq 
+\frac{{F_9}\,{r_{{i_1}}}\,{r_{{i_2}}}\,{v_{{i_3}}}\,
       (r \cdot v)\,(\epsilon {{\gamma }^{{i_1}}}\theta )\,
       (\theta {{\gamma }^{{i_2}{i_3}}}\theta )}{r^{11}}+
    \frac{{F_{10}}\,{r_{{i_1}}}\,{v_{{i_2}}}\,{v_{{i_3}}}\,
       (\epsilon {{\gamma }^{{a_1}{i_1}{i_2}}}\theta )\,
       (\theta {{\gamma }^{{a_1}{i_3}}}\theta )}{r^9} \brkeq 
+    \frac{{F_{11}}\,{r_{{i_1}}}\,{v_{{i_2}}}\,{v_{{i_3}}}\,
       (\epsilon {{\gamma }^{{i_2}}}\theta )\,
       (\theta {{\gamma }^{{i_1}{i_3}}}\theta )}{r^9} 
    +\frac{{F_{12}}\,{r_{{i_1}}}\,(r \cdot v)\,
       (\epsilon {{\gamma }^{{a_1}{a_2}{i_1}}}{\dot{\theta}}
        )\,(\theta {{\gamma }^{{a_1}{a_2}}}\theta )}{r^9}+
    \frac{{F_{13}}\,{v_{{i_1}}}\,
       (\epsilon {{\gamma }^{{a_1}{a_2}{i_1}}}{\dot{\theta}}
        )\,(\theta {{\gamma }^{{a_1}{a_2}}}\theta )}{r^7} 
\brkeq  +
    \frac{{F_{14}}\,{r_{{i_1}}}\,(r \cdot v)\,
       (\epsilon {{\gamma }^{{a_1}}}{\dot{\theta}})\,
       (\theta {{\gamma }^{{a_1}{i_1}}}\theta )}{r^9} 
    +\frac{{F_{15}}\,{r_{{i_1}}}\,{r_{{i_2}}}\,{v_{{i_3}}}\,
       (\epsilon {{\gamma }^{{a_1}{i_1}{i_3}}}{\dot{\theta}}
        )\,(\theta {{\gamma }^{{a_1}{i_2}}}\theta )}{r^9}+
    \frac{{F_{16}}\,{v_{{i_1}}}\,
       (\epsilon {{\gamma }^{{a_1}}}{\dot{\theta}})\,
       (\theta {{\gamma }^{{a_1}{i_1}}}\theta )}{r^7}
\brkeq +
    \frac{{F_{17}}\,{r_{{i_1}}}\,{r_{{i_2}}}\,{v_{{i_3}}}\,
       (\epsilon {{\gamma }^{{i_1}}}{\dot{\theta}})\,
       (\theta {{\gamma }^{{i_2}{i_3}}}\theta )}{r^9} 
    +\frac{{F_{18}}\,{r_{{i_1}}}\,(r \cdot v)\,
       (\epsilon {{\gamma }^{{a_1}{a_2}{a_3}{i_1}}}
        {\dot{\theta}})\,
       (\theta {{\gamma }^{{a_1}{a_2}{a_3}}}\theta )}{r^9}
\brkeq  +
    \frac{{F_{19}}\,{v_{{i_1}}}\,
       (\epsilon {{\gamma }^{{a_1}{a_2}{a_3}{i_1}}}
        {\dot{\theta}})\,
       (\theta {{\gamma }^{{a_1}{a_2}{a_3}}}\theta )}{r^7}
+
    \frac{{F_{20}}\,{r_{{i_1}}}\,(r \cdot v)\,
       (\epsilon {{\gamma }^{{a_1}{a_2}}}{\dot{\theta}})\,
       (\theta {{\gamma }^{{a_1}{a_2}{i_1}}}\theta )}{r^9}
    \brkeq  +\frac{{F_{21}}\,{r_{{i_1}}}\,{r_{{i_2}}}\,
       {v_{{i_3}}}\,(\epsilon 
        {{\gamma }^{{a_1}{a_2}{i_1}{i_3}}}{\dot{\theta}})\,
       (\theta {{\gamma }^{{a_1}{a_2}{i_2}}}\theta )}{r^9}+
    \frac{{F_{22}}\,{v_{{i_1}}}\,
       (\epsilon {{\gamma }^{{a_1}{a_2}}}{\dot{\theta}})\,
       (\theta {{\gamma }^{{a_1}{a_2}{i_1}}}\theta )}{r^7}
     \brkeq+
    \frac{{F_{23}}\,{r_{{i_1}}}\,{r_{{i_2}}}\,{v_{{i_3}}}\,
       (\epsilon {{\gamma }^{{a_1}{i_1}}}{\dot{\theta}})\,
       (\theta {{\gamma }^{{a_1}{i_2}{i_3}}}\theta )}{r^9}
+\frac{{F_{24}}\,{r_{{i_1}}}\,
       (\epsilon {{\gamma }^{{a_1}{a_2}{i_1}}}
        {\ddot{\theta}})\,
       (\theta {{\gamma }^{{a_1}{a_2}}}\theta )}{r^7}
+
    \frac{{F_{25}}\,{r_{{i_1}}}\,
       (\epsilon {{\gamma }^{{a_1}}}{\ddot{\theta}})\,
       (\theta {{\gamma }^{{a_1}{i_1}}}\theta )}{r^7}
    \brkeq +
    \frac{{F_{26}}\,{r_{{i_1}}}\,
       (\epsilon {{\gamma }^{{a_1}{a_2}{a_3}{i_1}}}
        {\ddot{\theta}})\,
       (\theta {{\gamma }^{{a_1}{a_2}{a_3}}}\theta )}{r^7}
+\frac{{F_{27}}\,{r_{{i_1}}}\,
       (\epsilon {{\gamma }^{{a_1}{a_2}}}{\ddot{\theta}})\,
       (\theta {{\gamma }^{{a_1}{a_2}{i_1}}}\theta )}{r^7}+
    \frac{{F_{28}}\,{r_{{i_1}}}\,
       (\epsilon {{\gamma }^{{a_1}{a_2}{i_1}}}\theta )\,
       ({\dot{\theta}}{{\gamma }^{{a_1}{a_2}}}{\dot{\theta}}
        )}{r^7}
    \brkeq  +\frac{{F_{29}}\,{r_{{i_1}}}\,
       (\epsilon {{\gamma }^{{a_1}}}\theta )\,
       ({\dot{\theta}}{{\gamma }^{{a_1}{i_1}}}{\dot{\theta}}
        )}{r^7} 
+
    \frac{{F_{30}}\,{r_{{i_1}}}\,
       (\epsilon {{\gamma }^{{a_1}{a_2}{a_3}{i_1}}}\theta 
        )\,({\dot{\theta}}{{\gamma }^{{a_1}{a_2}{a_3}}}
        {\dot{\theta}})}{r^7}+
    \frac{{F_{31}}\,{r_{{i_1}}}\,
       (\epsilon {{\gamma }^{{a_1}{a_2}}}\theta )\,
       ({\dot{\theta}}{{\gamma }^{{a_1}{a_2}{i_1}}}
        {\dot{\theta}})}{r^7}\period
\label{totepth3}
\end{align}
%
\shead{Construction of $X_{\ep\theta^6}$:}\quad 
The number of possible structures is 99, which is rather 
non-trivial to generate. However, after proper use of Fierz identities,
this gets drastically reduced to the following expression consisting of 
just 4 terms with arbitrary constants $G_i$:
\begin{align}
X_{\epsilon\theta^7} =  \and  \frac{{G_1}\,{r_{{i_1}}}\,
       (\epsilon {{\gamma }^{{a_1}{a_2}}}\theta )\,
       (\theta {{\gamma }^{{a_2}{a_3}}}\theta )\,
       (\theta {{\gamma }^{{a_3}{a_4}}}\theta )\,
       (\theta {{\gamma }^{{a_1}{a_4}{i_1}}}\theta )}{r^
       {11}}  \nn \\ \and 
 +\frac{{G_2}\,{r_{{i_1}}}\,{r_{{i_2}}}\,
       {r_{{i_3}}}\,(\epsilon {{\gamma }^{{a_1}}}\theta )\,
       (\theta {{\gamma }^{{a_1}{i_1}}}\theta )\,
       (\theta {{\gamma }^{{a_2}{i_2}}}\theta )\,
       (\theta {{\gamma }^{{a_2}{i_3}}}\theta )}{r^{13}}
 \nn \\ \and 
  +\frac{{G_3}\,{r_{{i_1}}}\,{r_{{i_2}}}\,
       {r_{{i_3}}}\,(\epsilon {{\gamma }^{{a_1}{i_3}}}
        \theta )\,(\theta {{\gamma }^{{a_2}{a_3}}}\theta )\,
       (\theta {{\gamma }^{{a_3}{i_2}}}\theta )\,
       (\theta {{\gamma }^{{a_1}{a_2}{i_1}}}\theta )}{r^
       {13}} \nn \\ \and    +\frac{{G_4}\,{r_{{i_1}}}\,{r_{{i_2}}}\,
       {r_{{i_3}}}\,(\epsilon {{\gamma }^{{a_1}{a_2}}}
        \theta )\,(\theta {{\gamma }^{{a_2}{i_3}}}\theta )\,
       (\theta {{\gamma }^{{a_3}{i_2}}}\theta )\,
       (\theta {{\gamma }^{{a_1}{a_3}{i_1}}}\theta )}{r^
       {13}}\period
\label{totepth7}
\end{align}
This completes the construction of the most general $X$ at order 3.

\subsection{Solution of the Ward identity}
We are now ready to solve the Ward identities (\ref{treesusy0}) $\sim$
 (\ref{treesusy8}). Although in the end we will find that all the unknown 
coefficients vanish, we describe the consequence of each of these
 equations separately. 

First consider (\ref{treesusy0}). Since $X_{\ep\del^3\theta}=0$ as 
 we argued, we only need to take the appropriate SUSY variations
 of $\Delta\calL^{(1)}_{\theta^0}$ and $\Delta\calL^{(1)}_{\theta^2}$, 
 given in (\ref{eff44}) and  (\ref{eff43}), remove the total derivatives 
 and set the result to zero. 
Collecting the structures of the same type, we get the equation
\begin{align}
\and 0=   \delta_r^{(0)}\calL^{(1)}_{\theta^0} + 
\delta_\theta^{(0)} \calL^{(1)}_{\theta^2} =  -
 \frac{\left( 14\,{B_5} + {B_6} - 7\,{B_9}
 - 
       2\,{B_{10}} \right)i \,{r_k}\,{v_i}\,{a_j}\,
    (r \cdot v)\,(\epsilon {{\gamma }^{ijk}}\theta )}{r^9} 
\brkeq 
- \frac{\left( 2\,{A_6} + 2\,{B_1} \right)i \,
     {{{\ddot{a}}}_i}\,(\epsilon {{\gamma }^{i}}\theta )}
     {r^5} - \frac{\left( 2\,{A_4} - 4\,{A_5} - 10\,{A_6} - 
       5\,{B_1} - 2\,{B_4} - {B_9} - 2\,{B_{11}} \right)i \,
     {a_i}\,v^2\,(\epsilon {{\gamma }^{i}}\theta )}{r^7} 
\brkeq  - 
  \frac{\left( -2\,{A_3} - 7\,{A_4} - 7\,{A_5} + {B_6} - 
       2\,{B_8} \right)i  \,{r_i}\,v^4\,
     (\epsilon {{\gamma }^{i}}\theta )}{r^9} - 
  \frac{\left( 2\,{A_2} + 2\,{A_4} - 5\,{A_6} + 
       2\,{B_5} \right)i \,{r_i}\,a^2\,
     (\epsilon {{\gamma }^{i}}\theta )}{r^7} 
 \brkeq - 
  \frac{\left( -2\,{A_3} + 70\,{A_6} + 35\,{B_1} - 
       2\,{B_2} - {B_6} + 7\,{B_9} + 2\,{B_{10}} \right)i \,
     {a_i}\,{(r \cdot v)}^2\,
     (\epsilon {{\gamma }^{i}}\theta )}{r^9} 
\brkeq  - 
  \frac{\left( -12\,{A_1} + 9\,{A_3} + 63\,{A_4} - 
       9\,{B_6} - 2\,{B_7} \right)i \,{r_i}\,v^2\,
     {(r \cdot v)}^2\,(\epsilon {{\gamma }^{i}}\theta )}
     {r^{11}} - \frac{33i\,{A_1}\,{r_i}\,{(r \cdot v)}^4\,
     (\epsilon {{\gamma }^{i}}\theta )}{r^{13}} 
\brkeq -
  \frac{21\,i{A_2}\,{r_i}\,{(r \cdot a)}^2\,
     (\epsilon {{\gamma }^{i}}\theta )}{r^9} - 
  \frac{\left( -20\,{A_6} - 15\,{B_1} - {B_9} \right)i \,
     {{{\dot{a}}}_i}\,(r \cdot v)\,
     (\epsilon {{\gamma }^{i}}\theta )}{r^7} \brkeq  - 
  \frac{\left( -4\,{A_3} - 14\,{A_4} + 28\,{A_5} - 
       2\,{B_2} + 7\,{B_4} - {B_6} + 2\,{B_8} \right) i \,
     {v_i}\,v^2\,(r \cdot v)\,
     (\epsilon {{\gamma }^{i}}\theta )}{r^9}  \brkeq  -
  \frac{\left( 18\,{A_3} + 9\,{B_2} + 9\,{B_6} + 
       2\,{B_7} \right) i \,{v_i}\,{(r \cdot v)}^3\,
     (\epsilon {{\gamma }^{i}}\theta )}{r^{11}}
 -
  \frac{\left( 4\,{A_2} + 2\,{A_4} + 2\,{B_5} + {B_9}
       \right) i \,{r_i}\,(v \cdot {\dot{a}})\,
     (\epsilon {{\gamma }^{i}}\theta )}{r^7}
 \brkeq  - 
  \frac{\left( 4\,{A_2} - 2\,{A_4} - 10\,{A_6} - 5\,{B_1} - 
       2\,{B_3} - 2\,{B_5} \right)i \,{a_i}\,(r \cdot a)\,
     (\epsilon {{\gamma }^{i}}\theta )}{r^7} 
-
  \frac{\left( -2\,{B_5} + {B_9} \right) i \,{r_k}\,{v_i}\,
     {{{\dot{a}}}_j}\,(\epsilon {{\gamma }^{ijk}}\theta )}
     {r^7} \brkeq  - 
  \frac{\left( -14\,{A_2} - 2\,{A_3} - 14\,{A_4} + 
       {B_6} \right)i \,{r_i}\,v^2\,(r \cdot a)\,
     (\epsilon {{\gamma }^{i}}\theta )}{r^9} - 
  \frac{\left( -12\,{A_1} + 126\,{A_2} \right)i \,{r_i}\,
     {(r \cdot v)}^2\,(r \cdot a)\,
     (\epsilon {{\gamma }^{i}}\theta )}{r^{11}} \brkeq  - 
  \frac{\left( -28\,{A_2} - 4\,{A_3} + 14\,{A_4} - 
       2\,{B_2} + 7\,{B_3} + 14\,{B_5} - {B_6} \right)i \,
     {v_i}\,(r \cdot v)\,(r \cdot a)\,
     (\epsilon {{\gamma }^{i}}\theta )}{r^9}\brkeq  - 
  \frac{\left( 4\,{A_2} - 2\,{A_4} - {B_3} - 2\,{B_5}
       \right)i \,{v_i}\,(r \cdot {\dot{a}})\,
     (\epsilon {{\gamma }^{i}}\theta )}{r^7} + 
  \frac{28i\,{A_2}\,{r_i}\,(r \cdot v)\,(r \cdot {\dot{a}})\,
     (\epsilon {{\gamma }^{i}}\theta )}{r^9} \brkeq
-   \frac{2i\,{A_2}\,{r_i}\,(r \cdot {\ddot{a}})\,
     (\epsilon {{\gamma }^{i}}\theta )}{r^7} 
  - 
  \frac{\left( 4\,{A_2} + 2\,{A_4} - 8\,{A_5} - {B_3} - 
       2\,{B_4} + {B_9} + 2\,{B_{11}} \right) i \,{v_i}\,
     (v \cdot a)\,(\epsilon {{\gamma }^{i}}\theta )}{r^7} \brkeq  
- 
  \frac{\left( -28\,{A_2} - 4\,{A_3} - 28\,{A_4} - 
       14\,{B_5} + {B_6} - 7\,{B_9} - 2\,{B_{10}} \right)i \,
     {r_i}\,(r \cdot v)\,(v \cdot a)\,
     (\epsilon {{\gamma }^{i}}\theta )}{r^9} .
\end{align}
This produces 23 relations for 17 unknowns, which appears to be an 
 overdetermined system. However, the solutions do exist and we obtain 
 the following 15 relations:
\begin{align}
 \and  {A_1} = 0, \quad {A_2} = 0, \ \ {A_3} = -\frac{7}{2}\,{A_4}, \quad 
   {A_6} = \frac{A_4}{5}, \brkeq  {B_1} = \frac{{A_4}}{5},  \quad 
   {B_2} = \frac{-7\,{A_4}}{2}, \quad  {B_3} = {A_4}, \quad 
   {B_4} = -{A_4} + 3\,{A_5}, \quad  {B_5} = \frac{{A_4}}{2}, \quad 
   {B_6} = -7\,{A_4}, \brkeq {B_7} = \frac{63\,{A_4}}{4}, \quad  
   {B_8} = \frac{-7\,\left( {A_4} - {A_5} \right) }{2}, \quad 
   {B_9} = {A_4},\ \ {B_{10}} = \frac{-7\,{A_4}}{2}, \quad 
   {B_{11}} = -{A_5}. \label{sol1}
\end{align}

Next we analyze the second Ward identity (\ref{treesusy2}). 
Starting at this order, a large number of terms are produced after 
 taking the relevant SUSY variations on the LHS, which are extremely 
 difficult to fully simplify into independent structures by 
 Fierz identities. Our strategy is to perform such simplification 
 as much as possible and then use an explicit representation of 
 the $SO(9)$ $\ga$-matrices to identify the coefficients of 
every independent structure written in terms of the components 
 of $r^m$, $\theta^\alpha$ and their derivatives. In this way, the Ward 
 identity is turned into a large set of coupled linear equations for 
 the coefficients. Solving them with the aid of Mathematica, we found 
 that all the coefficients involved in (\ref{treesusy2}) vanish, 
namely, 
\begin{align}
B_i = 0 \, , (i=1 \sim 11) \, , \quad
C_j = 0 \, , (j=1 \sim 13) \, , \quad 
F_k = 0 \, , (k=1 \sim 31) \, . 
\end{align}
Together with the previous result (\ref{sol1}), this implies that 
 $A_i$ must all vanish as well.

The third Ward identity (\ref{treesusy4}) was analyzed in 
 an entirely similar fashion. The result is that again all the 
 relevant coefficients must vanish:
\begin{eqnarray}
C_i = 0 \, , (i=1 \sim 13) \, , \quad D_j=0 \, , (j=1\sim 4) \period
\end{eqnarray}

Likewise,  the 4th Ward identity (\ref{treesusy6}) turned out to 
 dictate 
\begin{eqnarray}
D_i=0 \, , (i=1\sim 4) \, , \quad E_j=0 \, , (j=1\sim 3) \, ,
\quad G_k=0 \, , (k=1\sim 4 )\period
\end{eqnarray}
Finally, consider the last Ward identity (\ref{treesusy8}). 
This is in fact the same as the relation studied in \cite{Pabanetal1}.
Our result agrees with \cite{Pabanetal1} and gave the following 
 relations among the coefficients $E_i$:
\begin{align}
 E_1 = \frac{2}{143}E_3, \ \ E_2 = \frac{4}{13}E_3.
\end{align}
Since $E_i$'s must vanish by the previous equation, this is just a 
 consistency check. 

Putting the results of all the analyses together, we found a 
remarkable fact 
\begin{align}
 \Delta\Ga^{(1)} = 0. 
\end{align}
The precise meaning of this result is as follows: At 1-loop at order 4 
 in the derivative expansion, the solution to the Ward 
 identity $\delta_\ep \Ga=0$ regarded as a functional differential 
 equation for $\Ga$ is unique, provided that it exists. 
We would like to emphasize that, although we must assume the 
 existence of the solution (which of course is guaranteed in Matrix theory),
 our demonstration does not make use 
 of the explicit form of the quantum-corrected $\delta_\ep^{(1)}$ nor 
 the knowledge of $\Ga^{(1)}$: Only the 
 structure of the tree-level SUSY transformation law is relevant. 
We shall discuss the implication of this result further in the final section. 
\section{Discussions}
Since the conclusion of our present study is extremely simple and clear, 
 we shall in the following discuss its implications and make some 
 remarks. 

The fundamental question that we have been trying to answer in 
 the series of investigations is how much of the structures of 
Matrix theory is governed by its symmetries, in particular by the 
supersymmetry. This requires the study of off-shell effective action $\Ga$
 as well, since at the quantum level the very notion of \lq\lq on-shell" is 
defined by the full quantum-corrected equations of motion generated by $\Ga$. 
What we have found in this work is that once we assume the existence
 of the quantum corrected SUSY transformation operator $\delta_\ep$, 
 then the SUSY Ward identity, together with $SO(9)$ and CPT symmetries, 
dictates that, as far as the structures that would appear at 
 1-loop at order 4 in the system described by $r^m(\tau)$ and 
$\theta^\al(\tau)$ are concerned, fixes the off-shell effective action 
 uniquely.  The significance of our demonstration is that it does not
 refer to the underlying Matrix theory. This we believe is an important 
 step in understanding the genuine power of maximal supersymmetry. 
As we have already mentioned in the introduction,  our demonstration at
 order 4 is quite non-trivial since we know that from this order 
 the real dynamics starts, namely that the S-matrix is non-trivial. 

Would this property persist at higher orders? The fact that our logic
 was based solely on the tree-level SUSY transformation laws and that 
 our result at order 4 is highly unlikely to be just an accident makes us 
 feel that the answer ought to be yes. In fact we have already developed 
 an analytic scheme by which such a proof appears feasible. 
 The details will be presented in a forthcoming paper\cite{Kaz-Mura4}. 

Finally, we wish to make a remark on the meaning of the real 
power of supersymmetry. Although we have demonstrated the uniqueness
 of the solution of the SUSY Ward identity, it is not enough to claim 
 that SUSY determines the dynamics. Obviously we have not given 
 prescriptions to compute $\delta_\ep$ and $\Ga$ without referring 
 to Matrix theory.  If it is at all possible to prove such a claim,
 perhaps under 
 some favorable conditions,  one must study the closure property 
 of the effective SUSY transformations, which inevitably gets intertwined 
 with the knowledge of $\Ga$ itself. In other words, $\delta_\ep$ and $\Ga$
must be determined simultaneously in a self-consistent manner. 
Investigation of this problem is left for future study. 
\par\bigskip\noindent
{\large\bf Acknowledgment}\par\smallskip\noindent
We thank H. Nicolai for clarifying comments.
The research of Y.K. is supported in part by 
Grant-in-Aid for Scientific Research
on Priority Area \#707 \lq\lq Supersymmetry and Unified Theory of 
 Elementary Particles" No.~10209204 and 
 Grant-in-Aid for Scientific Research (B) 
No.~12440060, while that of T.M. is supported in part by 
the Japan Society for Promotion of Science under the Predoctoral
Research Program No.~12-9617, 
all from the Japan Ministry of Education, Culture, Sports, Science
 and Technology. 
\newpage
\setcounter{equation}{0}
\renewcommand{\theequation}{A.\arabic{equation}}
\noindent
{\Large\bf Appendix:} {\large\bf \quad
List of  {\boldmath $SO(9)$} Fierz identities}
\parbigskipn
We list the $SO(9)$ Fierz identities used to simplify 
 various expressions. They are presented in the order they are 
 referred to in the text and  classified by the number of 
 uncontracted indices called \lq\lq free indices". 
\shead{Identities used at $\calO(\theta^4)$:}

\begin{itemize}
\item 0-free-index type:
\begin{align}
  (\theta {{\gamma }^{{a_1}{a_2}}}\theta )(\theta 
     {{\gamma }^{{a_1}{a_2}}}\theta )  & = 0 , \ \ 
    (\theta {{\gamma }^{{a_1}{a_2}{a_3}}}\theta )(
     \theta {{\gamma }^{{a_1}{a_2}{a_3}}}\theta )  =  0 \period \label{F01}\\
 (\theta {{\gamma }^{{a_1}{a_2}}}\theta )\,
     ({\dot{\theta}}{{\gamma }^{{a_1}{a_2}}}\theta ) &  = 
     0, \ \  (\theta {{\gamma }^{{a_1}{a_2}{a_3}}}
      \theta )\,({\dot{\theta}}{{\gamma }^{{a_1}{a_2}{a_3}}}
      \theta )  = 0 \period \label{F02}
\end{align}
\item 2-free-index type with an $\ep$-tensor:
\begin{align}
(\theta {{\gamma }^{{m_1}{m_2}{m_3}}}\theta )\,
   ({\dot{\theta}}{{\gamma }^{{a_1}{a_2}{a_3}{a_4}}}\theta 
    )\,{{\epsilon }_{{a_1}{a_2}{a_3}{a_4}ij{m_1}{m_2}{m_3}}}
   = 720({\dot{\theta}}\theta )\,
    (\theta {{\gamma }^{ij}}\theta ) \period \label{F2ep1}
\end{align}
\item {Another 0-free-index type}
\begin{align}
 ({\dot{\theta}}{{\gamma }^{{a_1}{a_2}}}\theta )(
   {\dot{\theta}}{{\gamma }^{{a_1}{a_2}}}\theta ) = 
   6({\dot{\theta}}\theta )({\dot{\theta}}\theta )
    +2({\dot{\theta}}{{\gamma }^{{a_1}}}\theta )(
      {\dot{\theta}}{{\gamma }^{{a_1}}}\theta ) . \label{F03}
\end{align}
\item { 2-free-index type} 
\begin{align}
 (\theta {{\gamma }^{{a_1}{a_2}i}}\theta )\,
   (\theta {{\gamma }^{{a_1}{a_2}j}}\theta ) = &
  2(\theta {{\gamma }^{{a_1}i}}\theta )\,
    (\theta {{\gamma }^{{a_1}j}}\theta ) \, , \label{F20} \\
(\theta {{\gamma }^{{a_1}{a_2}i}}\theta )\,
     ({\dot{\theta}}{{\gamma }^{{a_1}{a_2}j}}\theta )   = &  
 8({\dot{\theta}}\theta )\,
       (\theta {{\gamma }^{ij}}\theta )+2
      (\theta {{\gamma }^{{a_1}j}}\theta )\,
      ({\dot{\theta}}{{\gamma }^{{a_1}i}}\theta ), \label{F21}\\
   (\theta {{\gamma }^{{a_1}ij}}\theta )\,
     ({\dot{\theta}}{{\gamma }^{{a_1}}}\theta )  = &  
      (\theta {{\gamma }^{{a_1}i}}\theta )\,
       ({\dot{\theta}}{{\gamma }^{{a_1}j}}\theta )
 -3({\dot{\theta}}\theta )\,
       (\theta {{\gamma }^{ij}}\theta )-
      (\theta {{\gamma }^{{a_1}j}}\theta )\,
       ({\dot{\theta}}{{\gamma }^{{a_1}i}}\theta )
, \label{F22} \\
(\theta {{\gamma }^{{a_1}{a_2}}}\theta )\,
     ({\dot{\theta}}{{\gamma }^{{a_1}{a_2}ij}}\theta )   = &  
  2({\dot{\theta}}\theta )\,
       (\theta {{\gamma }^{ij}}\theta )-2
      (\theta {{\gamma }^{{a_1}j}}\theta )\,
       ({\dot{\theta}}{{\gamma }^{{a_1}i}}\theta )+2
      (\theta {{\gamma }^{{a_1}i}}\theta )\,
       ({\dot{\theta}}{{\gamma }^{{a_1}j}}\theta ) , \label{F23} \\
 (\theta {{\gamma }^{{a_1}{a_2}j}}\theta )\,
   ({\dot{\theta}}{{\gamma }^{{a_1}{a_2}i}}{\dot{\theta}}) =
    \and -16({\dot{\theta}}{{\gamma }^i}\theta )\,
      ({\dot{\theta}}{{\gamma }^j}\theta )+2
     (\theta {{\gamma }^{{a_1}i}}\theta )\,
      ({\dot{\theta}}{{\gamma }^{{a_1}j}}{\dot{\theta}})-16
     ({\dot{\theta}}\theta )\,
      ({\dot{\theta}}{{\gamma }^{ij}}\theta ) \nn \\ & 
    -4({\dot{\theta}}\theta )({\dot{\theta}}\theta )\,
      {{\delta }_{ij}}
+  4  ({\dot{\theta}}{{\gamma }^{{a_1}}}\theta )(
       {\dot{\theta}}{{\gamma }^{{a_1}}}\theta )\,
      {{\delta }_{ij}} . \label{24}
\end{align}
\item { 4-free-index type:}
\begin{align}
 (\theta {{\gamma }^{{a_1}ij}}\theta )\,
   (\theta {{\gamma }^{{a_1}kl}}\theta ) = 
   \and  2(\theta {{\gamma }^{il}}\theta )\,
      (\theta {{\gamma }^{jk}}\theta )-2
     (\theta {{\gamma }^{ik}}\theta )\,
      (\theta {{\gamma }^{jl}}\theta )-3
     (\theta {{\gamma }^{ij}}\theta )\,
      (\theta {{\gamma }^{kl}}\theta ) \brkeq 
    +(\theta {{\gamma }^{{a_1}j}}\theta )\,
      (\theta {{\gamma }^{{a_1}l}}\theta )\,{{\delta }_{ik}}
      -(\theta {{\gamma }^{{a_1}j}}\theta )\,
      (\theta {{\gamma }^{{a_1}k}}\theta )\,{{\delta }_{il}}      \brkeq 
      -(\theta {{\gamma }^{{a_1}i}}\theta )\,
      (\theta {{\gamma }^{{a_1}l}}\theta )\,{{\delta }_{jk}}
+(\theta {{\gamma }^{{a_1}i}}\theta )\,
      (\theta {{\gamma }^{{a_1}k}}\theta )\,{{\delta }_{jl}} . \label{F41}
\end{align}
\end{itemize}
\pagebreak[2]
\shead{Identities used at $\calO(\theta^6)$:}
\begin{itemize}
 \item { 3-free-index type with an $\ep$-tensor:}
\begin{align}
 (\dot{\theta} {{\gamma }^{{a_1}{a_2}{a_3}}}\theta 
      )\,\and(\theta {{\gamma }^{{m_1}{m_2}{m_3}}}\theta )\,
     {{\epsilon }_{{a_1}{a_2}{a_3}ijk{m_1}{m_2}{m_3}}}  = 
     36(\dot{\theta} {{\gamma }^{{a_1}ij}}\theta )\,
       (\theta {{\gamma }^{{a_1}k}}\theta )+144
      (\dot{\theta} {{\gamma }^j}\theta )\,
       (\theta {{\gamma }^{ik}}\theta ) \brkeq 
     -144(\dot{\theta} {{\gamma }^i}\theta )\,
       (\theta {{\gamma }^{jk}}\theta )-36
      (\dot{\theta} {{\gamma }^{{a_1}k}}\theta )\,
       (\theta {{\gamma }^{{a_1}ij}}\theta )
     +18(\dot{\theta} {{\gamma }^{{a_1}{a_2}}}\theta )\,
       (\theta {{\gamma }^{{a_1}{a_2}j}}\theta )\,
       {{\delta }_{ik}}  \brkeq 
-18 (\dot{\theta} {{\gamma }^{{a_1}{a_2}}}\theta )\,
       (\theta {{\gamma }^{{a_1}{a_2}i}}\theta )\,
       {{\delta }_{jk}}, \label{F3ep2} \landfz 
 (\dot{\theta} {{\gamma }^{{a_1}{a_2}{a_3}{a_4}}}
      \theta )\,\and(\theta {{\gamma }^{{m_1}{m_2}}}\theta )\,
     {{\epsilon }_{{a_1}{a_2}{a_3}{a_4}ijk{m_1}{m_2}}}  = 
     72(\dot{\theta} {{\gamma }^{{a_1}ij}}\theta )\,
       (\theta {{\gamma }^{{a_1}k}}\theta )+240
      (\dot{\theta} {{\gamma }^k}\theta )\,
       (\theta {{\gamma }^{ij}}\theta ) \brkeq 
     +48(\dot{\theta} {{\gamma }^j}\theta )\,
       (\theta {{\gamma }^{ik}}\theta )-48
      (\dot{\theta} {{\gamma }^i}\theta )\,
       (\theta {{\gamma }^{jk}}\theta ) 
     -72(\dot{\theta} {{\gamma }^{{a_1}k}}\theta )\,
       (\theta {{\gamma }^{{a_1}ij}}\theta ) \brkeq  +36
      (\dot{\theta} {{\gamma }^{{a_1}{a_2}}}\theta )\,
       (\theta {{\gamma }^{{a_1}{a_2}j}}\theta )\,
       {{\delta }_{ik}} 
     -36(\dot{\theta} {{\gamma }^{{a_1}{a_2}}}\theta )\,
       (\theta {{\gamma }^{{a_1}{a_2}i}}\theta )\,
       {{\delta }_{jk}} , \label{F3ep3} \landfz
 (\theta {{\gamma }^{{a_1}{a_2}{a_3}}}\theta )\,
   \and(\theta {{\gamma }^{{m_1}{m_2}{m_3}}}\theta )\,
   {{\epsilon }_{{a_1}{a_2}{a_3}ijk{m_1}{m_2}{m_3}}} =   0 .
\label{F3ep4}
\end{align}
 \item { 4-free-index type with an $\ep$-tensor:}
\begin{align}
 (\theta {{\gamma }^{{m_1}{m_2}}}\theta )\,
   (\theta {{\gamma }^{{a_1}{a_2}{a_3}}}\theta )\,
   {{\epsilon }_{{a_1}{a_2}{a_3}ijkl{m_1}{m_2}}} = 
   \and -24(\theta {{\gamma }^{il}}\theta )\,
      (\theta {{\gamma }^{jk}}\theta )+24
     (\theta {{\gamma }^{ik}}\theta )\,
      (\theta {{\gamma }^{jl}}\theta ) \brkeq 
    -24(\theta {{\gamma }^{ij}}\theta )\,
      (\theta {{\gamma }^{kl}}\theta ) . \label{F4ep1}
\end{align}
\item 5-free-index type with an $\ep$-tensor:
\begin{align} (\theta {{\gamma }^{{a_1}{a_2}i}}\theta )\,\and
   (\theta {{\gamma }^{{m_1}{m_2}{m_3}}}\theta )\,
   {{\epsilon }_{{a_1}{a_2}jklm{m_1}{m_2}{m_3}}}  
=  \brkeq 
    -12(\theta {{\gamma }^{lm}}\theta )\,
      (\theta {{\gamma }^{ijk}}\theta )+12
     (\theta {{\gamma }^{km}}\theta )\,
      (\theta {{\gamma }^{ijl}}\theta )-12
     (\theta {{\gamma }^{kl}}\theta )\,
      (\theta {{\gamma }^{ijm}}\theta ) \brkeq 
    -12(\theta {{\gamma }^{jm}}\theta )\,
      (\theta {{\gamma }^{ikl}}\theta )+12
     (\theta {{\gamma }^{jl}}\theta )\,
      (\theta {{\gamma }^{ikm}}\theta )-12
     (\theta {{\gamma }^{jk}}\theta )\,
      (\theta {{\gamma }^{ilm}}\theta ) \brkeq 
    -12(\theta {{\gamma }^{im}}\theta )\,
      (\theta {{\gamma }^{jkl}}\theta )+12
     (\theta {{\gamma }^{il}}\theta )\,
      (\theta {{\gamma }^{jkm}}\theta )-12
     (\theta {{\gamma }^{ik}}\theta )\,
      (\theta {{\gamma }^{jlm}}\theta ) \brkeq 
    +12(\theta {{\gamma }^{ij}}\theta )\,
      (\theta {{\gamma }^{klm}}\theta ) \label{F3ep1}
\end{align}
\end{itemize}
\begin{itemize}
\item 1-free-index type:
\begin{align}
  (\theta {{\gamma }^{{a_1}{a_2} i }}\theta )(\theta 
     {{\gamma }^{{a_1}{a_2}}}\theta ) = 0 \period \label{F00}
\end{align}
\item 3-free-index type:
\begin{align}
 (\theta {{\gamma }^{{a_1}{a_2}k}}\theta )\,
   ({\dot{\theta}}{{\gamma }^{{a_1}{a_2}ij}}\theta )\and = 
    -8(\theta {{\gamma }^{jk}}\theta )\,
      ({\dot{\theta}}{{\gamma }^i}\theta )+8
     (\theta {{\gamma }^{ik}}\theta )\,
      ({\dot{\theta}}{{\gamma }^j}\theta )+10
     (\theta {{\gamma }^{ij}}\theta )\,
      ({\dot{\theta}}{{\gamma }^k}\theta ) \brkeq 
    -2(\theta {{\gamma }^{{a_1}ij}}\theta )\,
      ({\dot{\theta}}{{\gamma }^{{a_1}k}}\theta )+2
     (\theta {{\gamma }^{{a_1}k}}\theta )\,
      ({\dot{\theta}}{{\gamma }^{{a_1}ij}}\theta ) \brkeq  +2
     (\theta {{\gamma }^{{a_1}{a_2}j}}\theta )\,
      ({\dot{\theta}}{{\gamma }^{{a_1}{a_2}}}\theta )\,
      {{\delta }_{ik}} 
    -2(\theta {{\gamma }^{{a_1}{a_2}i}}\theta )\,
      ({\dot{\theta}}{{\gamma }^{{a_1}{a_2}}}\theta )\,
      {{\delta }_{jk}} \period \label{F32}
\end{align}
\item 4-free-index type:
\begin{align}
(\theta {{\gamma }^{{a_1}ij}}\theta )\,
   ({\dot{\theta}}{{\gamma }^{{a_1}kl}}\theta ) = 
   \and -(\theta {{\gamma }^{jkl}}\theta )\,
      ({\dot{\theta}}{{\gamma }^i}\theta )+
     (\theta {{\gamma }^{ikl}}\theta )\,
      ({\dot{\theta}}{{\gamma }^j}\theta )+
     (\theta {{\gamma }^{ijl}}\theta )\,
      ({\dot{\theta}}{{\gamma }^k}\theta ) \brkeq 
    -(\theta {{\gamma }^{ijk}}\theta )\,
      ({\dot{\theta}}{{\gamma }^l}\theta )-2
     (\theta {{\gamma }^{kl}}\theta )\,
      ({\dot{\theta}}{{\gamma }^{ij}}\theta )-
     (\theta {{\gamma }^{jl}}\theta )\,
      ({\dot{\theta}}{{\gamma }^{ik}}\theta ) \brkeq 
    +(\theta {{\gamma }^{jk}}\theta )\,
      ({\dot{\theta}}{{\gamma }^{il}}\theta )+
     (\theta {{\gamma }^{il}}\theta )\,
      ({\dot{\theta}}{{\gamma }^{jk}}\theta )-
     (\theta {{\gamma }^{ik}}\theta )\,
      ({\dot{\theta}}{{\gamma }^{jl}}\theta ) \brkeq 
    -(\theta {{\gamma }^{ij}}\theta )\,
      ({\dot{\theta}}{{\gamma }^{kl}}\theta )-
     ({\dot{\theta}}\theta )\,
      (\theta {{\gamma }^{jl}}\theta )\,{{\delta }_{ik}}-
     (\theta {{\gamma }^{{a_1}jl}}\theta )\,
      ({\dot{\theta}}{{\gamma }^{{a_1}}}\theta )\,
      {{\delta }_{ik}} \brkeq 
    +(\theta {{\gamma }^{{a_1}j}}\theta )\,
      ({\dot{\theta}}{{\gamma }^{{a_1}l}}\theta )\,
      {{\delta }_{ik}}+({\dot{\theta}}\theta )\,
      (\theta {{\gamma }^{jk}}\theta )\,{{\delta }_{il}}+
     (\theta {{\gamma }^{{a_1}jk}}\theta )\,
      ({\dot{\theta}}{{\gamma }^{{a_1}}}\theta )\,
      {{\delta }_{il}} \brkeq 
    -(\theta {{\gamma }^{{a_1}j}}\theta )\,
      ({\dot{\theta}}{{\gamma }^{{a_1}k}}\theta )\,
      {{\delta }_{il}}+({\dot{\theta}}\theta )\,
      (\theta {{\gamma }^{il}}\theta )\,{{\delta }_{jk}}+
     (\theta {{\gamma }^{{a_1}il}}\theta )\,
      ({\dot{\theta}}{{\gamma }^{{a_1}}}\theta )\,
      {{\delta }_{jk}} \brkeq 
    -(\theta {{\gamma }^{{a_1}i}}\theta )\,
      ({\dot{\theta}}{{\gamma }^{{a_1}l}}\theta )\,
      {{\delta }_{jk}}-({\dot{\theta}}\theta )\,
      (\theta {{\gamma }^{ik}}\theta )\,{{\delta }_{jl}}-
     (\theta {{\gamma }^{{a_1}ik}}\theta )\,
      ({\dot{\theta}}{{\gamma }^{{a_1}}}\theta )\,
      {{\delta }_{jl}} \brkeq 
    +(\theta {{\gamma }^{{a_1}i}}\theta )\,
      ({\dot{\theta}}{{\gamma }^{{a_1}k}}\theta )\,
      {{\delta }_{jl}} \period \label{F42}
\end{align}
\end{itemize}
\shead{Identities used for total derivative terms:}
\begin{itemize}
\item { 1-free-index type:}
\begin{align}
(\epsilon {{\gamma }^{{a_1}{a_2}}}\theta )\,
     (\theta {{\gamma }^{{a_1}{a_2}i}}\theta )  = 
    \and -2(\epsilon {{\gamma }^{{a_1}}}\theta )\,
      (\theta {{\gamma }^{{a_1}i}}\theta ), \label{F12}
   \landfz 
(\epsilon {{\gamma }^{{a_1}{a_2}i}}\theta )\,
     (\theta {{\gamma }^{{a_1}{a_2}}}\theta )  = 
    \and +2(\epsilon {{\gamma }^{{a_1}}}\theta )\,
      (\theta {{\gamma }^{{a_1}i}}\theta ), \label{F13}
   \landfz 
(\epsilon {{\gamma }^{{a_1}{a_2}{a_3}i}}\theta 
      )\,(\theta {{\gamma }^{{a_1}{a_2}{a_3}}}\theta )  = 
    \and -6(\epsilon {{\gamma }^{{a_1}}}\theta )\,
      (\theta {{\gamma }^{{a_1}i}}\theta ) . \label{F14}
\end{align}
\item {3-free-index type:} 
\begin{align}
 (\epsilon {{\gamma }^{{a_1}{a_2}ij}}\theta )\,
   (\theta {{\gamma }^{{a_1}{a_2}k}}\theta )  = \and  
 2(\epsilon {{\gamma }^{{a_1}ij}}\theta )\,
      (\theta {{\gamma }^{{a_1}k}}\theta )+10
     (\epsilon {{\gamma }^k}\theta )\,
      (\theta {{\gamma }^{ij}}\theta )+8
     (\epsilon {{\gamma }^j}\theta )\,
      (\theta {{\gamma }^{ik}}\theta ) \brkeq 
    -8(\epsilon {{\gamma }^i}\theta )\,
      (\theta {{\gamma }^{jk}}\theta )-2
     (\epsilon {{\gamma }^{{a_1}k}}\theta )\,
      (\theta {{\gamma }^{{a_1}ij}}\theta )-4
     (\epsilon {{\gamma }^{{a_1}}}\theta )\,
      (\theta {{\gamma }^{{a_1}j}}\theta )\,{{\delta }_{ik}}
     \brkeq +4(\epsilon {{\gamma }^{{a_1}}}\theta )\,
      (\theta {{\gamma }^{{a_1}i}}\theta )\,{{\delta }_{jk}} , 
\label{F33} \\[0.6ex]
 (\epsilon {{\gamma }^{{a_1}k}}\theta )\,
   (\theta {{\gamma }^{{a_1}ij}}\theta ) = \and  
   -(\epsilon {{\gamma }^{{a_1}jk}}\theta )\,
      (\theta {{\gamma }^{{a_1}i}}\theta )+
     (\epsilon {{\gamma }^{{a_1}ik}}\theta )\,
      (\theta {{\gamma }^{{a_1}j}}\theta )+3
     (\epsilon {{\gamma }^k}\theta )\,
      (\theta {{\gamma }^{ij}}\theta ) \brkeq 
    +(\epsilon {{\gamma }^j}\theta )\,
      (\theta {{\gamma }^{ik}}\theta )-
     (\epsilon {{\gamma }^i}\theta )\,
      (\theta {{\gamma }^{jk}}\theta )+
     (\epsilon \theta )\,(\theta {{\gamma }^{ijk}}\theta )
     \brkeq -(\epsilon {{\gamma }^{{a_1}}}\theta )\,
      (\theta {{\gamma }^{{a_1}j}}\theta )\,{{\delta }_{ik}}
      +(\epsilon {{\gamma }^{{a_1}}}\theta )\,
      (\theta {{\gamma }^{{a_1}i}}\theta )\,{{\delta }_{jk}} \period 
\label{F34}
\end{align}
\end{itemize}

\newpage


\begin{thebibliography}
\bibitem{Kaz-Mura1} Y. Kazama and T. Muramatsu, Nucl. Phys. {\bf 584}
 (2000) 171, hep-th/0003161.
\bibitem{Kaz-Mura2} Y. Kazama and T. Muramatsu, Class.\ Quant.\ Grav.\
	{\bf 18}, 2277 (2001), hep-th/0103116.
\bibitem{bfss} T. Banks, W. Fischler, S. H. Shenker and L. Susskind,
Phys. Rev. {\bf D55} (1997) 5112, hep-th/9610043.
\bibitem{susskind} L. Susskind, hep-th/9704080.
\bibitem{halpernetal} For pioneering works, see 
M.~Claudson and M.~B.~Halpern, Nucl.\ Phys.\ B {\bf 250}, 689 (1985);
R.~Flume, Annals Phys.\  {\bf 164}, 189 (1985);
M.~Baake, M.~Reinicke and V.~Rittenberg, 
J.\ Math.\ Phys.\  {\bf 26}, 1070 (1985); 
B.~de Wit, J.~Hoppe and H.~Nicolai, Nucl.\ Phys.\ B {\bf 305}, 545 (1988).
\bibitem{reviews}
T.~Banks, Nucl.\ Phys.\ Proc.\ Suppl.\  {\bf 67}, 180 (1998), hep-th/9710231; \
T.~Banks, hep-th/9911068; \ 
D.~Bigatti and L.~Susskind, hep-th/9712072; \ 
A.~Bilal, Fortsch.\ Phys.\  {\bf 47}, 5 (1999), hep-th/9710136; \
B.~de Wit, hep-th/9902051; \ 
A.~Konechny and A.~Schwarz, hep-th/0012145; \ 
H.~Nicolai and R.~Helling, hep-th/9809103; \ 
N.~A.~Obers and B.~Pioline, Phys.\ Rept.\  {\bf 318}, 113 (1999), hep-th/9809039; \
W.~Taylor, hep-th/9801182; \
W.~Taylor, hep-th/0002016; \ 
W.~Taylor, hep-th/0101126.
\bibitem{seiberg} N. Seiberg, Phys. Rev. Lett. {\bf 79} (1997) 3577,
 hep-th/9710009.
\bibitem{sen} A. Sen,  Adv. Theor. Math. Phys. {\bf 2} (1998) 51,
 hep-th/9709220.
\bibitem{Pabanetal1} S. Paban, S. Sethi and M. Stern,
Nucl. Phys. {\bf B534} (1998) 137, hep-th/9805018.
\bibitem{Pabanetal2} S. Paban, S. Sethi and M. Stern,
J. High Energy Phys. {\bf 06} (1998) 012, hep-th/9806028.
\bibitem{lowe} D. A. Lowe, J. High Energy Phys. {\bf 11} (1998) 009, 
hep-th/9810075.
\bibitem{ss} S. Sethi and M. Stern,
J. High Energy Phys. {\bf 06} (1999) 004, hep-th/9903049.
\bibitem{Hyunetal} S. Hyun, Y. Kiem and H. Shin, Nucl. Phys. 
{\bf B558} (1999) 349, hep-th/9903022.
\bibitem{np} H. Nicolai and J. Plefka, Phys. Lett. {\bf B477} (2000) 309,
hep-th/0001106.
\bibitem{Kaz-Mura4} Y. Kazama and T. Muramatsu, work in progress. 
\end{thebibliography}
\end{document}